\documentclass[twocolumn,epjc3]{svjour3}

\usepackage{hyperref}
\usepackage{epsfig,graphics,colordvi}
\usepackage{amsmath}
\usepackage{amssymb}
\usepackage{mciteplus}
\usepackage{dsfont}
\usepackage{color}
\smartqed  % flush right qed marks, e.g. at end of proof
\RequirePackage{graphicx}
\journalname{Eur. Phys. J. C}
\begin{document}

\title{Photon-fusion reactions from the chiral Lagrangian with
dynamical light vector mesons}
%\titlerunning{Short form of title}        % if too long for running head
\author{I.V. Danilkin\thanksref{addr1,addr3}%{e1,addr1,addr3}
        \and
        M.F.M.\ Lutz\thanksref{addr1} %etc.
        \and
        S.\ Leupold\thanksref{addr2} %etc.
        \and
        C.\ Terschl\"usen\thanksref{addr2} %etc.
}
%\thankstext{e1}{e-mail: i.danilkin@gsi.de}
\institute{GSI Helmholtzzentrum f\"ur Schwerionenforschung GmbH,\\
Planckstra\ss e 1, 64291 Darmstadt, Germany\label{addr1}
           \and
Institutionen f\"or fysik och astronomi, Uppsala Universitet, Box
516, 75120 Uppsala, Sweden\label{addr2}
           \and
SSC RF ITEP, Bolshaya Cheremushkinskaya 25, 117218 Moscow,
Russia\label{addr3} }

%\date{Received: date / Accepted: date}
% The correct dates will be entered by the editor

\maketitle

\begin{abstract}
We study the reactions $\gamma\gamma\rightarrow \pi^0\pi^0$,
$\pi^+\pi^-$, $K^0\bar{K}^0$, $K^+K^-$, $\eta\,\eta$ and
$\pi^0\eta$ based on a chiral Lagrangian with dynamical light
vector mesons as formulated within the hadrogenesis conjecture. At
present our chiral Lagrangian contains 5 unknown parameters that
are relevant for the photon fusion reactions. They parameterize
the strength of interaction terms involving two vector meson
fields. These parameters are fitted to photon fusion data
$\gamma\gamma\rightarrow \pi^0\pi^0$, $\pi^+\pi^-, \pi^0\eta$  and
to the decay $\eta\rightarrow\pi^0\gamma\gamma$. In order to
derive gauge invariant reaction amplitudes in the resonance region
constraints from maximal analyticity and exac t coupled-channel
unitarity are used. Our results are in good agreement with the
existing experimental data from threshold up to about 0.9 GeV for
the two-pion final states. The $a_0$ meson in the $\pi^0\eta$
channel is dynamically generated and an accurate reproduction of
the $\gamma\gamma\rightarrow \pi^0\eta$ data is achieved up to 1.2
GeV. Based on our parameter sets we predict the
$\gamma\gamma\rightarrow $ $K^0\bar{K}^0$, $K^+K^-$, $\eta\,\eta$
cross sections. 
\end{abstract}

\section{Introduction}
\label{intro}

Photon-fusion reactions $\gamma\gamma\rightarrow PP$ (with
$PP=\pi^0\pi^0$, $\pi^+\pi^-$, $K^0\bar{K}^0$, $K^+K^-$,
$\eta\eta$ and $\pi^0\eta$) play an important role in our
understanding of non-perturbative QCD
\cite{Pennington:2008xd,Oller:2008kf,GarciaMartin:2010cw,Hoferichter:2011wk,Mao:2009cc}.
As a systematic approach chiral perturbation theory ($\chi$PT) is
applied to describe these reactions at low energies
\cite{Gasser:2006qa,Gasser:2005ud}. Such studies have been
performed at next-to-leading order (one loop)
\cite{Bijnens:1987dc,Donoghue:1988eea} and at
next-to-next-to-leading order (two loops)
\cite{Gasser:2005ud,Gasser:2006qa}. However, $\chi$PT is limited
to the near-threshold region and cannot serve as an appropriate
framework in the resonance region, where exact coupled-channel
unitarity becomes an important issue. In
\cite{Gasparyan:2010xz,Danilkin:2010xd,Danilkin:2011fz,Gasparyan:2011yw,Gasparyan:2012km}
it was shown that the combination of electromagnetic gauge
invariance, maximal analyticity and coupled-channel unitarity together
with chiral symmetry helps to achieve a systematic approximation
of hadron interactions beyond the threshold region. In the
following the same approach is applied to the photon-fusion
reactions.

The cross sections of fusion processes are very sensitive to
hadronic final-state interactions. Therefore, a crucial input for
the present work is a proper description of the reactions of two
pseudoscalars into the same or different two pseudoscalars.
Recently, within the novel unitarization approach developed in
\cite{Gasparyan:2010xz,Danilkin:2010xd,Gasparyan:2011yw,Gasparyan:2012km}
a coupled-channel computation of Goldstone-boson scattering has
been performed \cite{Danilkin:2011fz}. The calculations are based
on the chiral Lagrangian formulated with light vector mesons.

The light vector mesons play a crucial role in the hadrogenesis
conjecture
\cite{Lutz:2001dr,Lutz:2003fm,Lutz:2004dz,Lutz:2005ip,Lutz:2007sk,Lutz:2008km,Terschlusen:2012xw}.
Together with the Goldstone bosons they are identified to be the
``quasi-fundamental'' hadronic degrees of freedom that are
expected to generate the meson spectrum. For instance it was shown
that the leading chiral interaction of Goldstone bosons with the
light vector mesons generates an axial-vector meson spectrum that
is quite close to the empirical one \cite{Lutz:2003fm}. It seems
quite natural to keep vector mesons as explicit degrees of freedom
when aiming for a description of hadron physics in the resonance
region. In particular, this extends the resonance-saturation
mechanism \cite{Ecker:1988te} by taking into account explicitly
the dynamics of the vector-meson propagator \cite{Kubis:2000zd}.
In the present analysis we further emphasize the significant role
of light vector mesons in hadron physics.

One important aspect in photon-fusion reactions is the formation
of scalar and tensor resonances. In particular one finds the
rather narrow scalar states $a_0(980)$ and $f_0(980)$
\cite{Oller:1997yg,Pennington:2008xd,Mao:2009cc}. In the present
work we concentrate on energies below 1.2 GeV. There the scalar
resonances $f_0(980)$ and $a_0(980)$ appear to dominate. In the
approach of \cite{Danilkin:2011fz} both states are dynamically
generated from coupled-channel interactions, in accordance with
the hadrogenesis conjecture. The two-pion channel, where the
$f_0(980)$ resonance is seen prominently, has been discussed in
detail in \cite{Danilkin:2011fz}. In contrast, the $\pi\,\eta$
channel with the $a_0(980)$ state has not been addressed in
\cite{Danilkin:2011fz}. There are no elastic $\pi\,\eta$
scattering data available. However, the $\pi\,\eta$ channel can be
populated by the inelastic photon-fusion reaction. We will show in
the following that the dynamical generation of the $a_0(980)$ is
in very good agreement with the available experimental data on the
photon-fusion production of $\pi\,\eta$.

At higher energies tensor resonances come into play. In principle,
also the tensor resonances $f_2(1270)$ and $a_2(1320)$ are
expected to be naturally generated within our approach from
vector-vector interactions. However, we focus in the present paper
on lower energies and on the mutual interactions between Goldstone
bosons. In the present approach vector mesons appear as exchange
particles, but not as states in the coupled channels, i.e.\ we
consider scattering and rescattering of the type $ \gamma\gamma
\to PP$ and $PP \to PP$, respectively, but disregard $PP
\leftrightarrow VV, \, VP$, where $V$ generically denotes a
vector-meson state. In principle, the basis for the systematic
inclusion of vector mesons as coupled-channel states has been laid
out in \cite{Lutz:2011xc}. This technically more challenging
computation is beyond the scope of the present work. Consequently
the analysis of the resonances $f_2(1270)$ and $a_2(1320)$ and the
corresponding energy regime is postponed to the future.

Experimental information on the recations $\gamma\gamma\rightarrow
\,$hadrons is accessible in $e^+e^-$ collisions via the reaction
$e^+ + e^- \rightarrow e^+ + e^- + \,$hadrons
\cite{Boyer:1990vu,Behrend:1992hy,Marsiske:1990hx,Antreasyan:1985wx,Albrecht:1989re,Behrend:1988hw,Althoff:1985yh}.
Recently high-statistics data have been reported by the Belle
Collaboration
\cite{Uehara:2009cka,Mori:2007bu,Uehara:2009cf,Uehara:2010mq},
including the first measurement of $\eta\eta$ production. The
reaction $\gamma\gamma \to \pi^0\,\eta$
%, where the $a_0(980)$ appears as a prominent peak,
is linked to the decay $\eta\rightarrow \pi^0\gamma\gamma$ by
crossing symmetry. Recent analyses of this decay have been
performed at AGS \cite{Prakhov:1900zz,Prakhov:2008zz} and at MAMI
\cite{Prakhov:1900zz,Unverzagt:2009vm}. We will use the
corresponding integrated and differential information about this
decay to constrain our so-far unknown coupling constants
parameterizing the strength of interaction terms with two vector
meson fields.

The main purpose of this paper is to obtain a unified description
of the reactions $\gamma\gamma\rightarrow\pi^0\pi^0$,
$\pi^+\pi^-$, $K^0\bar{K}^0$, $K^+K^-$, $\eta\eta$ and $\pi^0\eta$
using the novel framework introduced in \cite{Danilkin:2011fz,Terschlusen:2012xw}.
The ultimate goal of our studies is to generalize the dispersive effective field theory framework
\cite{Gasparyan:2010xz,Danilkin:2010xd,Gasparyan:2011yw,Gasparyan:2012km} to the chiral Lagrangian with explicit vector meson fields. The present study extends the analysis of Goldstone-boson
scattering performed in \cite{Danilkin:2011fz}. In the next
section we will specify our Lagrangian and comment on the
determination of the corresponding coupling constants. Section
\ref{sec:pwa} presents the required decomposition into
partial-wave amplitudes. The unitary and causal summation scheme,
which is at the heart of the coupled-channel dynamics, is reviewed
in section \ref{sec:ccd}. Numerical results and comparisons to
data are given in section \ref{sec:numres} and a summary of our
work is provided in section \ref{sec:sum}.

\section{Chiral interaction Lagrangian including vector mesons}
\label{sec:chiint}

The study of the photon-fusion processes is performed in
application of the chiral Lagrangian with explicit vector mesons.
The leading order terms of this Lagrangian have been constructed
based on a formal power counting scheme
\cite{Terschlusen:2012xw,Lutz:2008km}. The counting rests on the
dynamical assumption of hadrogenesis and on large-$N_c$ arguments,
where $N_c$ denotes the number of colors. In previous successful
applications two- and three-body decays of vector mesons have been
presented in \cite{Lutz:2008km,Leupold:2008b,Terschluesen:2010ik}.
For photon-fusion reactions the relevant part of the leading-order
Lagrangian takes the simple form\footnote{Note that in
\cite{Danilkin:2011fz,Lutz:2008km} slightly different notations
were used. The relations between $e_V$, $g_D$, $g_F$, $h_P$ in
\cite{Danilkin:2011fz,Lutz:2008km} --- denoted by {\it old} ---
and $f_V$, $g_1$, $g_2$, $h_P$ used here and in
\cite{Terschlusen:2012xw} are $f_V= \frac{0.776 \, \rm
GeV}{4\,e}\,e_V$, $g_1=g_D$, $g_2=g_F$ and $h_P=\frac{0.776 \, \rm
GeV}{f_V}\,h_P[old]$.} {\allowdisplaybreaks
\begin{eqnarray}\label{Eq:LagrangianGGtoPP}
\mathcal{L}&=&-\frac{e^2}{2}\,A^{\mu}\,A_{\mu}\,\textrm{tr}\,\big\{\Phi\,Q\,\big[\Phi,\,Q\big]_-\big\}\nonumber\\
&&{}+i\,\frac{e}{2}\,A^\mu\,\textrm{tr}\,\big\{\partial_\mu\Phi\,\big[Q,\,\Phi\big]_-\big\}\nonumber\\
&&{}-e\,f_V\,\partial_{\mu}A_{\nu}\,\textrm{tr}\big\{\Phi^{\mu\nu}\,Q\big\}\nonumber\\
&&{}-i\,\frac{f_V\,h_P}{8\,f^2}\,\textrm{tr}\big\{\partial_{\mu}\Phi\,\Phi^{\mu\nu}\,\partial_{\nu}\Phi\big\}\nonumber\\
&&{}+\frac{e\,f_V}{8\,f^2}\,\partial_{\mu}A_{\nu}\,\textrm{tr}\big\{\Phi^{\mu\nu}\big[\Phi,\big[\Phi,\,Q\big]_{-}\big]_{-}\big\}\nonumber\\
&&{}+\frac{e\,f_V\,h_P}{8f^2}\,A_{\nu}\,{\rm tr}\big\{\big[\partial_\mu\Phi\,,\,\Phi^{\mu\nu}\big]_-\big[Q,\Phi\big]_{-}\big\}\\
&&{}-\frac{1}{16\,f^2}\,\textrm{tr}\big\{\partial^\mu\Phi_{\mu\alpha}\,\big[\big[\Phi,\partial_{\nu}\Phi\big]_-,\Phi^{\nu\alpha}\big]_-\big\}\nonumber\\
&&{}-\frac{b_D}{64\,f^2}\,\textrm{tr}\big\{\Phi^{\mu\nu}\,\Phi_{\mu\nu}\,\big[\Phi,\big[\Phi,\chi_0\big]_+\big]_+\big\}\nonumber\\
&&{}-\frac{g_1}{32\,f^2}\,\textrm{tr}\,\big\{\big[\Phi_{\mu\nu}\,,\partial_\alpha\Phi\big]_+\,\big[\partial^\alpha\Phi,\Phi^{\mu\nu}\big]_+\big\}\nonumber\\
&&{}-\frac{g_2}{32\,f^2}\,\,\textrm{tr}\,\big\{\big[\Phi_{\mu\nu}\,,\partial_\alpha\Phi \big]_-\,\big[\partial^\alpha\Phi,\Phi^{\mu \nu}\big]_-\big\}\nonumber\\
&&{}-\frac{g_3}{32\,f^2}\,\,\textrm{tr}\,\big\{\big[\,\partial_\mu\Phi\,,\partial^\nu\Phi\big]_+\,\big[\Phi_{\nu\tau}\,,\Phi^{\mu\tau}\big]_+\big\}\nonumber\\
&&{}-\frac{g_5}{32\,f^2}\,\,\textrm{tr}\,\big\{\big[\Phi^{\mu\tau},\partial_\mu\Phi\big]_-\,\big[\Phi_{\nu\tau}\,,\partial^\nu\Phi\big]_-\big\}\nonumber\\
&&{}-\frac{h_A}{16\,f}\,\epsilon_{\mu\nu\alpha\beta}\,\textrm{tr}\big\{\big[\Phi^{\mu\nu},\partial_{\tau}\Phi^{\tau\alpha}\big]_+\,\partial^{\beta}\Phi\big\}\nonumber\\
&&{}-\frac{b_A}{16\,f}\,\epsilon_{\mu\nu\alpha\beta}\,\textrm{tr}\big\{\big[\Phi^{\mu\nu},\Phi^{\alpha\beta}\big]_+\big[\chi_0,\Phi\big]_+\big\}\nonumber\\
&&{}-\frac{h_O}{16\,f}\,\epsilon_{\mu\nu\alpha\beta}\,\textrm{tr}\big\{\big[\partial^{\alpha}\Phi^{\mu\nu},\Phi^{\tau\beta}\big]_+\partial_{\tau}\Phi\big\}\nonumber
\end{eqnarray}}
where the Goldstone-boson field $\Phi$, the vector-meson field
$\Phi_{\mu\nu}$, the charge matrix $Q$, and the mass matrix
$\chi_0$ are normalized as follows:
{\allowdisplaybreaks\begin{eqnarray}
&&\Phi=\left(\begin{array}{ccc}
\pi^0+\frac{1}{\sqrt{3}}\,\eta &\sqrt{2}\,\pi^+&\sqrt{2}\,K^+\\
\sqrt{2}\,\pi^-&-\pi^0+\frac{1}{\sqrt{3}}\,\eta&\sqrt{2}\,K^0\\
\sqrt{2}\,K^- &\sqrt{2}\,\bar{K}^0&-\frac{2}{\sqrt{3}}\,\eta
\end{array}\right)\,,\nonumber\\
&&\Phi_{\mu \nu} = \left(\begin{array}{ccc}
\rho^0_{\mu \nu}+\omega_{\mu \nu} &\sqrt{2}\,\rho_{\mu \nu}^+&\sqrt{2}\,K_{\mu \nu}^+\\
\sqrt{2}\,\rho_{\mu \nu}^-&-\rho_{\mu \nu}^0+\omega_{\mu \nu}&\sqrt{2}\,K_{\mu \nu}^0\\
\sqrt{2}\,K_{\mu \nu}^- &\sqrt{2}\,\bar{K}_{\mu
\nu}^0&\sqrt{2}\,\phi_{\mu \nu}
\end{array}\right)\,,\nonumber\\
&&\chi_0=\left( \begin{array}{ccc} m_\pi^2 & 0 & 0\\ 0 & m_\pi^2&0\\
0 & 0 & 2 \,m_K^2 - m_\pi^2 \end{array} \right)\,,
\nonumber\\
&& Q = \left( \begin{array}{ccc} \frac{2}{3} & 0 & 0 \\
0 & -\frac{1}{3} & 0 \\
 0 & 0 & - \frac{1}{3}
\end{array}
\right)\,.
\end{eqnarray}}
Note that we assume perfect isospin symmetry throughout this work.
In (\ref{Eq:LagrangianGGtoPP}) the photon field is denoted by
$A_\mu$ and $e=0.303$ is the electromagnetic charge. For vector
mesons we use the antisymmetric tensor-field representation
$\Phi_{\mu\nu}=-\Phi_{\nu\mu}$ giving rise to the
resonance-saturation mechanism \cite{Ecker:1988te}. In principle,
in the hadrogenesis conjecture also the singlet eta field is part
of the ``quasi-fundamental'' hadronic degrees of freedom. It can
be included in the flavor matrix of $\Phi$ in a straightforward
way \cite{Terschlusen:2012xw}. However, as discussed in the
introduction, we do not include vector channels in our
coupled-channel approach. Consequently we also do not include
channels with the $\eta'$ which would appear in the same energy
regime. Vector mesons are important, nonetheless, as they
contribute as exchange particles to the coupled channels
$\gamma\gamma$ and $PP$. On the other hand, this is not the case
for the $\eta'$. Consequently, here we have not considered the eta
singlet explicitly in our Lagrangian.

Finally we shall discuss the coupling constants appearing in
(\ref{Eq:LagrangianGGtoPP}). The following set of parameters has
been determined by the masses and decay properties of the vector
mesons \cite{Lutz:2008km,Leupold:2008b}:
\begin{eqnarray}
\label{Eq:Set_of_parameters}
\begin{array}{ll}
f_V=0.140 \pm 0.014 ~ \textrm{GeV}\,, \qquad  \quad   &
h_A\simeq2.10\,,\\
h_P \, f_V = 0.23\,\mbox{GeV}\,,& b_D =
0.92\,, \\
f\simeq0.90~\textrm{GeV}\,, & b_A = 0.27\,.   \\
\end{array}
\end{eqnarray}
The values of the other parameters $g_{1-3}$, $g_5$ and $h_O$ have
not been determined so far. Assuming that they are of natural size
we will study in section \ref{sec:numres} the impact of variations
of these parameters on the photon-fusion processes and on the
related decay $\eta \to \pi^0\gamma\gamma$.

\begin{table}[t]
\centering \caption{\label{Tab:Isospin_states_GGtoPP}The
coupled-channel states $I^G$ characterized by isospin $I$ and
G-parity $G$. The Pauli matrices $\sigma_i$ act on isospin-doublet
fields $K,\bar{K}$ with for instance $K=(K^+,K^0)^t$. Note that in
particular the neutral ($I_3=0$) two-pion  state with isospin two
is given by
$\textstyle{\frac{1}{\sqrt{6}}}\,(2\,\pi^0_p\,\pi^0_q-\pi^+_p\,\pi^-_q-\pi^-_p\,\pi^+_q)$.}
%\tabcolsep=3.mm
\renewcommand{\arraystretch}{1.8}
%\fontsize{9}{11}
\begin{tabular*}{\columnwidth}{@{\extracolsep{\fill}}cc@{}}
\hline\noalign{\smallskip}
  $0^+$ & $1^-$ \\
\noalign{\smallskip}\hline\noalign{\smallskip}
$\left(\!\!\begin{array}{c}
(\gamma\,\gamma)\\
{\textstyle{1\over \sqrt{3}}}\,(
\pi_{q}\cdot\,\pi_{p}) \\
{\textstyle{1\over 2}}\,(\bar K_{q}\,K_{p}+\bar K_{p}\,K_{q})\\
(\eta_{q}\,\eta_{p})\\
\end{array}\!\!\right)$ &
$\left(\!\!\begin{array}{c}
(\gamma\,\gamma) \\
(\pi_{q}\,\eta_{p})_{I_3=0}\\
{\textstyle{1\over2}}\,(\bar{K}_{q}\,\vec\sigma\,K_{p}+\bar{K}_{p}\,\vec\sigma\,K_{q})_{I_3=0}\\
\end{array}\right)$
\\
\noalign{\smallskip}\hline\noalign{\smallskip}
\multicolumn{2}{c}{$2^+$}\\
\noalign{\smallskip}\hline\noalign{\smallskip}
\multicolumn{2}{@{}c}{$\left(\begin{array}{c}
(\gamma\,\gamma) \\
\left({\textstyle{1\over2}}\,(\pi^i_{q}\,\pi^j_{p}+\pi^j_{q}\,\pi^i_{p})-{\textstyle{1\over3}}\,
\delta_{ij}\,\pi_{q}\cdot\pi_{p}\right)_{I_3=0}\\
\end{array}\right)$}\\
\noalign{\smallskip}\hline
\end{tabular*}
\end{table}

\section{Partial-wave amplitudes}
\label{sec:pwa}

We define the transition amplitude for the process $\gamma\gamma
\rightarrow PP$ as
\begin{eqnarray}\label{eq:defTA}
&&\left<P(\bar{p}) \,P(\bar{q})|\,T\,|A(k_1,\lambda_1) \,
A(k_2,\lambda_2)\right>
= \\
&&\quad(2\pi)^4\delta^4(k_1+k_2-\bar{p}-\bar{q})
    \,\,T^{\mu\nu}\,\epsilon_{\mu}(k_1,\lambda_1)\,\epsilon_{\nu}(k_2,\lambda_2)\nonumber
\phantom{mm}
\end{eqnarray}
where $k_{1,2}$ and $\epsilon_{1,2}$ are the momenta and the
polarization vectors of the incoming photons, respectively, and
$\bar{p}$, $\bar{q}$ are the momenta of the outgoing mesons.

In general, the two-body scattering problem decouples into
orthogonal channels specified by isospin, G-parity, parity and
strangeness quantum numbers. For the case at hand, with two
photons in the initial state and two pseudoscalars in the final
state, parity is always positive (see below) and strangeness is
always zero. In each of the channels, finally specified by isospin
$I$ and G-parity $G$, there are several meson-meson states coupled
to each other. In Table \ref{Tab:Isospin_states_GGtoPP} we have
specified the states which contain the most relevant meson-meson
information below 1.2 GeV. Here we are neglecting multi-pion
states which are only relevant for higher energies. In particular
we neglect in this way also $PV$ and $VV$ states which on account
of the resonant nature of the vector, $V$, states would
significantly contribute to the multi-pion states.

\begin{figure*}[t]
\centering
\includegraphics[height=2cm,clip=true]{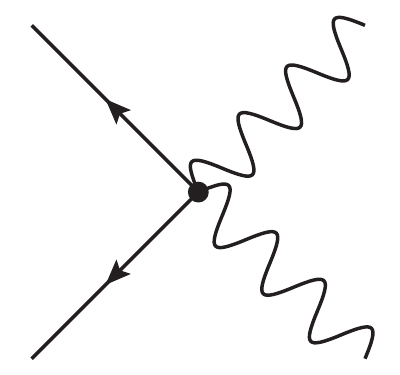}\quad\includegraphics[height=2cm,clip=true]{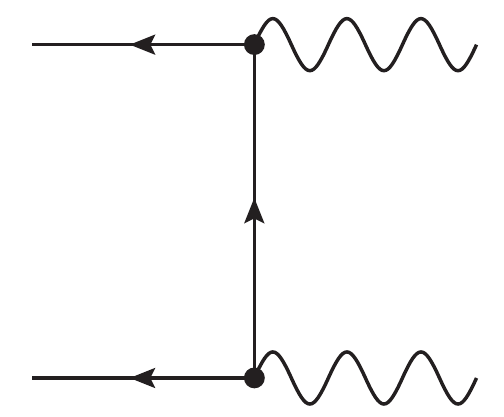}
\quad\includegraphics[height=2cm,clip=true]{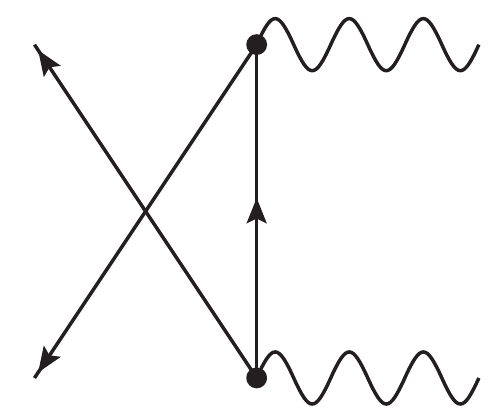}\quad\includegraphics[height=2cm,clip=true]{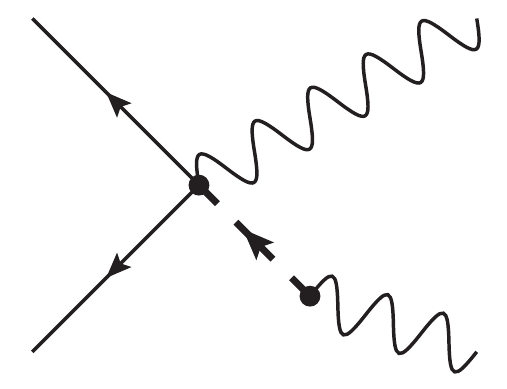}
\includegraphics[height=2cm,clip=true]{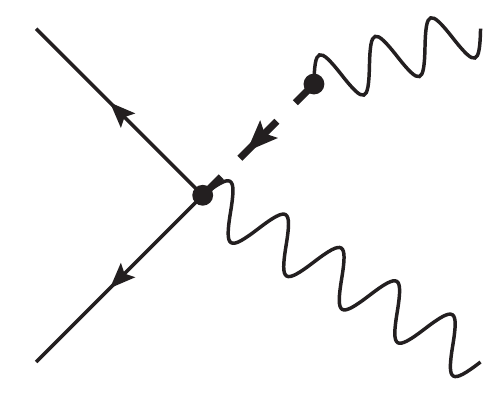}\quad\includegraphics[height=2cm,clip=true]{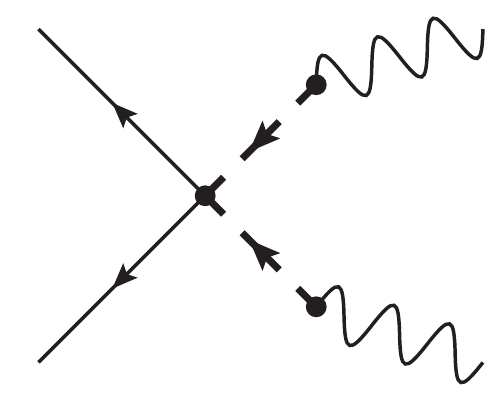}
\quad\includegraphics[height=2cm,clip=true]{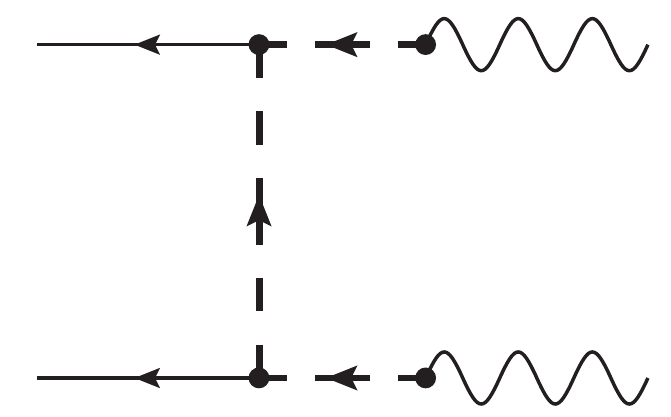}\quad\includegraphics[height=2cm,clip=true]{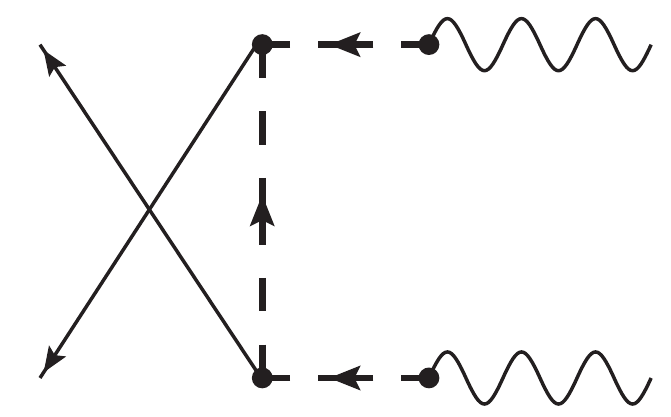}
\caption{\label{Fig:Tree_level_diagrams}Tree-level diagrams for
$\gamma\gamma\rightarrow PP$ reactions with the exchange of
pseudoscalar (solid line) and light vector (dashed line) mesons.}
\end{figure*}

Lorentz covariance and gauge invariance lead to a decomposition of
the scattering amplitude $~T^{\mu\nu}$ into Lorentz tensors
$L_i^{\mu\nu}$ and invariant amplitudes $F_{i}$,
\begin{eqnarray}\label{Eq:Lorez_structures}
T^{\mu\nu}&=&F_1(s,t,u)\,L_1^{\mu\nu}+F_2(s,t,u)\,L_2^{\mu\nu}\,,\nonumber\\[0.3em]
L_1^{\mu\nu}&=&k_1^{\nu}\,k_2^{\mu}-(k_1\cdot k_2)\,g^{\mu\nu}\,,\\
L_2^{\mu\nu}&=&(\Delta^2\,(k_1\cdot k_2)-2\,(k_1\cdot
\Delta)\,(k_2\cdot
\Delta))\,g^{\mu\nu}\nonumber\\
&&-\Delta^2\,k_1^\nu\,k_2^\mu
    -2(k_1\cdot k_2)\,\Delta^{\mu}\,\Delta^{\nu}\nonumber\\
    &&+2(k_2\cdot \Delta)\,k_1^{\nu}\,\Delta^{\mu}+2(k_1\cdot \Delta)\,k_2^{\mu}\,\Delta^{\nu}\,,\nonumber
\end{eqnarray}
where $\Delta=\bar{p}-\bar{q}$ and $T^{\mu\nu}$ satisfies the Ward
identities
\begin{equation}\label{Eq:Ward_identities}
    k_{1\mu}T^{\mu\nu}=k_{2\nu}T^{\mu\nu}=0\,.
\end{equation}

The motivation for choosing these particular Lorentz structures is
twofold. First, the corresponding invariant amplitudes are
independent and free of kinematical singularities or zeros.
Second, in order to simplify further calculations we have chosen
the Lorentz tensors such that the following property holds:
\begin{equation}\label{}
    L_1^{\mu\nu}\,L_{2\,\mu\nu}=0\,.
\end{equation}

The invariant amplitudes $F_1$ and $F_2$ are analytic functions of
$s$, $t$ and $u$ except for dynamical cuts. Furthermore, a $t-u$
crossing symmetry is satisfied due to the Bose statistics of the
two photons.

It is useful to introduce the helicity components of the
scattering amplitude and decompose each of them into their partial
waves,
%{\allowdisplaybreaks}
\begin{eqnarray}\label{Eq:Helicity_amplitudes}
\phi_{++}&=&T^{\mu\nu}\,\epsilon_\mu(k_1,+1)\,\epsilon_\nu(k_2,+1)\nonumber\\
&=&\sum_{\textrm{even }J\geq0}(2J+1)\,t^{(J)}_{++}\,d_{00}^{(J)}(\cos\theta)\,,\nonumber \\
\phi_{+-}&=&T^{\mu\nu}\,\epsilon_\mu(k_1,+1)\,\epsilon_\nu(k_2,-1)\nonumber\\
&=&\sum_{\textrm{even}J\geq2}(2J+1)\,
t^{(J)}_{+-}\,d_{20}^{(J)}(\cos\theta)\,,
\end{eqnarray}
where
\begin{eqnarray}
&&\epsilon_\mu(\vec{k}_1,\lambda=\pm1)=\left(%
\begin{array}{c}
  0 \\
  \mp\frac{1}{\sqrt{2}} \\
  -\frac{i}{\sqrt{2}} \\
  0 \\
\end{array}%
\right)\,,\nonumber\\
&&\epsilon_\nu(\vec{k}_2=-\vec{k}_1,\lambda=\pm1)=\left(%
\begin{array}{c}
  0 \\
  \pm\frac{1}{\sqrt{2}} \\
  -\frac{i}{\sqrt{2}} \\
  0 \\
\end{array}%
\right)\,, \nonumber
\end{eqnarray}
and $d_{\lambda,\bar{\lambda}}^{(J)}(\cos\theta)$ are Wigner
rotation functions. In (\ref{Eq:Helicity_amplitudes}) $\theta$ is
the center-of-mass scattering angle. The axis of the colliding
photons in their center-of-mass frame has been chosen to agree
with the third axis of the coordinate system.
%otherwise the explicit form of the epsilons does not make sense!!
Note that the partial-wave expansion involves only even $J\geq
\lambda$ and positive parity $P=+$ \cite{landau-qed}. This
constraint arises from the combination of Bose symmetry of the two
initial massless photons with the possible $J$, $P$ quantum
numbers of the two Goldstone bosons in the final state.

The invariant amplitudes $F_{1,2}$ can be expressed in terms of
the helicity amplitudes $\phi_{++}$, $\phi_{+-}$,
\begin{eqnarray}\label{}
    &&\left(%
\begin{array}{c}
  F_1 \\
  F_2 \\
\end{array}%
\right)=\left(
\begin{array}{cc}
 -\frac{2}{s} &  0 \\
 0 & \frac{1}{2\,s\,\bar{p}_{cm}^2\,(x^2-1)}
\end{array}
\right)\left(%
\begin{array}{c}
  \phi_{++} \\
  \phi_{+-} \\
\end{array}%
\right)
\end{eqnarray}
where $\bar{p}_{\rm cm}$ is the final center-of-mass relative
momentum and $x=\cos\theta$. For unpolarized photons the
differential cross section is given by
\begin{equation}\label{Eq:Cross_section}
    \frac{d\sigma}{d\cos\theta}=\frac{\beta}{32\,\pi\,s}\,\frac{1}{4}\,\left(2\,|\phi_{++}|^2+2\,|\phi_{+-}|^2\right)\,,
\end{equation}
where $\beta=2\,\bar{p}_{cm}/\sqrt{s}$. If two identical particles
appear in the final state (neutral pions, for instance) one has to
include an additional factor of $1/2$ in (\ref{Eq:Cross_section})
or perform the integration only over $\theta\in [0,\,\pi/2]$.

According to (\ref{Eq:Helicity_amplitudes}) the partial-wave
helicity amplitudes $t^{(J)}_{++}$, $t^{(J)}_{+-}$ can be computed
in terms of the invariant amplitudes $F_{1,2}$ as
\begin{eqnarray}
t^{(J)}_{++}(s)&=&-\int\frac{dx}{4}\,s\,F_1(s,x)\,d^{(J)}_{00}(x)\,,
\nonumber\\
t^{(J)}_{+-}(s)&=&\int\frac{dx}{2}\,2\,\bar{p}_{cm}^2\,s\,
\left(x^2-1\right)\,F_2(s,x)\,d^{(J)}_{20}(x) \,,
\label{Eq:P.w.amplitudes}
\end{eqnarray}
with the help of the useful identities for the Wigner rotation
functions \cite{Varshalovich1988}
\begin{eqnarray}\label{Eq:d-functions}
d_{00}^{(J)}(x)&=&P_J(x)\,,\nonumber\\
d_{20}^{(J)}(x)&=& \frac{2\,x\,P'_J(x)}{\sqrt{(J-1)\, J\, (J+1)\,
(J+2)}}\nonumber\\
&&\quad-\sqrt{\frac{J\, (J+1)}{(J-1)\, (J+2)}}\,P_J(x)\,.
\end{eqnarray}

In order to avoid kinematical singularities and zeros in the
partial-wave amplitudes at threshold, we rescale
(\ref{Eq:P.w.amplitudes}) by a phase-space factor
$(p_{cm}\,\bar{p}_{cm})^J$,
\begin{eqnarray}\label{Eq:P.w.amplitudes_new}
  T^{(J)}_{++} &=& \frac{s^J}{(p_{cm}\,\bar{p}_{cm})^J}\,t^{(J)}_{++}\,, \nonumber \\
  T^{(J)}_{+-} &=& \frac{s^J}{(p_{cm}\,\bar{p}_{cm})^J}\,t^{(J)}_{+-}\,,
\end{eqnarray}
and also multiply by $s^J$ to ensure a finite limit of the
phase-space matrices at large energy.

The invariant amplitudes (\ref{Eq:Lorez_structures}) computed from
the chiral Lagrangian (\ref{Eq:LagrangianGGtoPP}) read

{\allowdisplaybreaks%
\begin{eqnarray}
  F_1 &=& \frac{e^2\,C_{SG}}{2\,s}-\sum_{x\in[8]}\frac{e^2\,(m_x^4-t\,u)\,C_{0}^{(x)}}{2s^2\,(t-m_x^2)}\nonumber\\
  && {}+
  \sum_{x\in[9]}\left(\frac{e^2\,f_V^2}{8\,f^2}\,\frac{C_{1}^{(x)}}{m_x^2}
    -\frac{e^2\,f_V^2\,h_P}{8\,f^2}\,\frac{C_{h_P}^{(x)}}{m_x^2}\right)
  \nonumber\\
  && {} +\sum_{x,y\in[9]}\frac{e^2\,f_V^2}{32\,f^2}\,\frac{1}{m_x^2\,m_y^2}\,
  \Bigg( \left(g_1\,C_{g_1}^{(x,y)}+g_2\,C_{g_2}^{(x,y)} \right.\nonumber\\
  && {} \qquad \left. + \,g_3\,C_{g_3}^{(x,y)}+g_5\,C_{g_5}^{(x,y)}\right)(s-\bar{m}_1^2-\bar{m}_2^2)
  \nonumber\\
  &&\hspace*{5em} {}  -\frac{1}{2}\,b_D\,C_{b_D}^{(x,y)} \Bigg)
  \nonumber\\
  && {} +\sum\limits_{x,y,z\in[9]}
  \frac{e^2\,f_V^2}{16\,f^2}\,\frac{1}{m_x^2\,m_y^2}\, \Bigg( \frac{1}{16}\,\frac{t^2}{m_z^2}\,h_O^2\,C_{h_O}^{(x,y,z)}
   \nonumber\\
  &&\hspace*{5em} {}  -\frac{1}{16}\,\frac{t^2}{t-m_z^2}\,h_A^2\,C_{h_A}^{(x,y,z)}
  \nonumber\\
  &&\hspace*{5em} {} +
   \frac{t-2\,m_z^2}{m_z^2}\,
  \frac{1}{t-m_z^2}\, b_A^2\, C_{b_A}^{(x,y,z)}
  \nonumber\\
  &&\hspace*{5em} {}+\frac{1}{2}\,\frac{t}{t-m_z^2 }\,b_A\,h_A\,C_{b_A \,h_A}^{(x,y,z)}
  \nonumber\\
  &&\hspace*{5em}  {}
    +\frac{1}{2}\frac{t}{m_z^2}\,b_A\,h_O\,C_{b_A\,h_O}^{(x,y,z)}\Bigg)
  \nonumber\\
  && {} +(t\leftrightarrow u)  \,,
\nonumber \\
  F_2 &=& -\sum_{x\in[8]}\,\frac{e^2\,C_{0}^{(x)}}{8\,s\,(t-m_x^2)}
  \nonumber\\
  &&{}-\sum_{x,y\in[9]}\frac{e^2\,f_V^2}{64\,f^2}\,\frac{1}{m_x^2\,m_y^2}\,\left(g_3\,C_{g_3}^{(x,y)}+g_5\,C_{g_5}^{(x,y)}\right)
  \nonumber\\
  &&{} +\sum\limits_{x,y,z\in[9]}\frac{e^2\,f_V^2}{64\,f^2}\,\frac{1}{m_x^2\,m_y^2}\, \Bigg(
  \frac{1}{16}\,\frac{t}{m_z^2}\,h_O^2\,C_{h_O}^{(x,y,z)}
   \nonumber\\
  &&\hspace*{5em} {} +\frac{1}{16}\,\frac{t}{t-m_z^2}\,h_A^2\,C_{h_A}^{(x,y,z)}
  \nonumber\\
  &&\hspace*{5em} {}+\frac{1}{m_z^2}\,
   \frac{1}{t-m_z^2} \, b_A^2\,C_{b_A}^{(x,y,z)}
  \nonumber\\
  &&\hspace*{5em} {} -\frac{1}{2}\,
  \frac{1}{t-m_z^2 }\,b_A\,h_A\,C_{b_A\,h_A}^{(x,y,z)}
  \nonumber\\
  &&\hspace*{5em} {} +\frac{1}{m_z^2}\,b_A\,h_O\,C_{b_A\,h_O}^{(x,y,z)}\Bigg)
  \nonumber\\
  && {} +(t\leftrightarrow u) \,,
  \label{Eq:F1F2}
\end{eqnarray}}
\noindent where the sum runs over the octet of Goldstone bosons
([8]) or the vector-meson nonet ([9]) and $m_{x,y,z}$ denotes
their respective masses. The coefficients $C_{...}$ are presented
in Tables \ref{Tab:C[SG]}, \ref{Tab:C[bAhA]}, \ref{Tab:C[bA,hA]},
and \ref{Tab:C[g1,g2,g3,g5]} with respect to the coupled-channel
states of Table \ref{Tab:Isospin_states_GGtoPP}. In Fig.\
\ref{Fig:Tree_level_diagrams} the set of tree-level diagrams that
gives nonzero contributions is depicted.  Note that for the
isospin states which contain identical particles (e.g.
$|\pi\,\pi\rangle$, $|\eta\,\eta\rangle$, \ldots) we use a
convention where the unitarity condition for identical and
non-identical two-particle states are the same.

The conventions of Table \ref{Tab:Isospin_states_GGtoPP} imply the
following relations between scattering amplitudes in isospin and
particle bases,
\begin{eqnarray}
&&T_{\gamma\gamma\rightarrow\pi^+\pi^-}=2\,\Big(\frac{1}{\sqrt{3}}\,T^{I=0}_{\gamma\gamma\rightarrow\pi\,\pi}-\frac{1}{\sqrt{6}}\,T^{I=2}_{\gamma\gamma\rightarrow\pi\,\pi}\Big)\,,\nonumber\\
&&T_{\gamma\gamma\rightarrow\pi^0\pi^0}=2\,\Big(\frac{1}{\sqrt{3}}\,T^{I=0}_{\gamma\gamma\rightarrow\pi\,\pi}+\sqrt{\frac{2}{3}}\,T^{I=2}_{\gamma\gamma\rightarrow\pi\,\pi}\Big)\,,\nonumber\\
&&T_{\gamma\gamma\rightarrow\pi^0\eta}=\sqrt{2}\,\,T^{I=1}_{\gamma\gamma\rightarrow\pi\,\eta}\,,\qquad T_{\gamma\gamma\rightarrow\,\eta\,\eta}=2\,T^{I=0}_{\gamma\gamma\rightarrow\,\eta\,\eta}\,,\nonumber\\
&&T_{\gamma\gamma\rightarrow\,K^+K^-}=T^{I=0}_{\gamma\gamma\rightarrow\,K\,\bar{K}}+T^{I=1}_{\gamma\gamma\rightarrow\,K\,\bar{K}}\,,\nonumber\\
&&T_{\gamma\gamma\rightarrow\,K^0\bar{K^0}}=T^{I=0}_{\gamma\gamma\rightarrow\,K\,\bar{K}}-T^{I=1}_{\gamma\gamma\rightarrow\,K\,\bar{K}}\,,
\end{eqnarray}
where the factor 2 reflects our normalization for two-body states
with identical particles.  Note that in \cite{Danilkin:2011fz}
meson-meson interactions were studied only for S- and P-waves. The
result for the D-wave amplitude can easily be obtained from Eq.\
(4) and Eq.\ (5) of \cite{Danilkin:2011fz}.

The partial-wave amplitudes obtained from
(\ref{Eq:P.w.amplitudes_new}, \ref{Eq:P.w.amplitudes}) and
(\ref{Eq:F1F2}) at tree-level will serve as an input for the
non-perturbative coupled-channel calculations to which we turn
next.

\begin{figure}[t]
\centering
\includegraphics[keepaspectratio,width=0.5\textwidth]{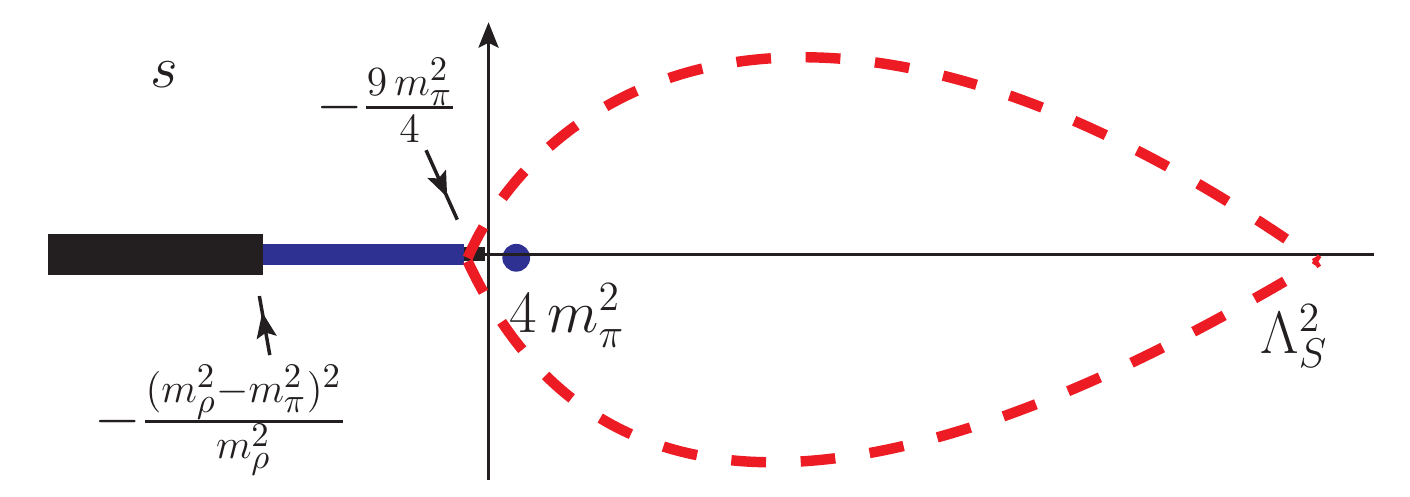}
\caption{Locations of left-hand cuts of the
$\gamma\gamma\rightarrow \pi\pi$ partial-wave amplitude in the
complex $s$-plane. The branch point of the $\rho$-meson exchange
is located at $\Lambda_0^2=-(m_{\rho}^2-m_{\pi}^2)^2/m_{\rho}^2$,
while one- and two-pion exchange cuts start at $\Lambda_0^2=0$,
$-\,9\,m_{\pi}^2/4$, respectively. The dashed line identifies the
convergence region of the conformal expansion
(\ref{Eq:U_expansion2}).\label{Fig:Mapping_GGPP}}
\end{figure}

\section{Coupled-channel dynamics}
\label{sec:ccd}

We assume that the analytic properties of the scattering amplitude, required to derive
partial-wave  dispersion relations, arise as a consequence of the
underlying micro-causality of local quantum field theory.  The well-known Mandelstam  postulate or also phrased as the
maximal analyticity assumption expects the scattering amplitude to be an analytic function everywhere except
for possible poles and cuts coming from the unitarity and crossing
symmetry constraints \cite{Mandelstam:1958xc,Mandelstam:1959bc}.
This expectation is confirmed in perturbation theory but it is difficult to prove
in a non-perturbative framework. For the $\pi\pi$ scattering it can be
rigorously derived in a finite region \cite{Martin:1965jj,Martin:1966jj}. However for photon-fusion
reactions the corresponding analyticity was not proven nor disproven so far.
Within the maximal analyticity assumption the  partial-wave reaction amplitudes satisfy the
dispersion-integral representation,
\begin{eqnarray}\label{Eq:Dispersion_relation}
T^J_{ab}(s)&=&U^J_{ab}(s)\\
&+&\sum_{c,d}\int^{\infty}_{\mu_{\rm
thr}^2}\frac{d\bar{s}}{\pi}\frac{s-\mu_M^2}{\bar{s}-\mu_M^2}\,
\frac{T^J_{ac}(\bar s)\,\rho^J_{cd}(\bar s)\,T^{J*}_{db}(\bar
s)}{\bar{s}-s-i\epsilon}\,,\nonumber
\end{eqnarray}
where the phase-space matrix $\rho^J_{cd}(s)$ is diagonal in $c$
and $d$. In (\ref{Eq:Dispersion_relation}) the coupled-channel
indices  $a$ and $b$  run over the various channel $\gamma\gamma$,
$\pi\pi$, $K \bar K$ etc. Crossing symmetry leads to a non-trivial coupling of
various partial-wave amplitudes $ T^J_{ab}(s)$ as is systematically exploited in the Roy-Steiner equations \cite{Roy:1971tc,Ananthanarayan:2000ht,Buettiker:2003pp,GarciaMartin:2011jx,Moussallam:2011zg,Hoferichter:2011wk}.
We return to this issue at the end of this section. 
The unitarity condition implies that the
discontinuity of the generalized potential
\begin{eqnarray}\label{Eq:Unitarity}
    \Delta U^{J}_{ab}(s)&=&\frac{1}{2\,i}\left(U_{ab}^J(s+i\epsilon)-U_{ab}^J(s-i\epsilon)\right)
    \nonumber\\
    &=& 0 \qquad {\rm for} \qquad s \geq \mu^2_{\rm thres} \ ,
\end{eqnarray}
vanishes for energies larger than the s-channel threshold. This is
nothing but the condition that  $U^{J}_{ab}(s)$ has left-hand cuts
only.

In our normalization the hadronic part of the phase-space matrix
\begin{equation}\label{Eq:Rho}
\rho^J (s)=
    \frac{1}{8 \pi } \left(\frac{p_{cm}}{\sqrt{s}}\right)^{2\,J+1}\Theta(s-\mu_{\textrm{thr}}^2)\,,
\end{equation}
approaches a finite value in the high-energy limit. Intermediate
states with two photons are neglected in this work. Numerically
they are largely suppressed being proportional to $e^4$ at least.
Therefore the two-dimensional phase-space matrix for the
two-photon states need not to be specified here.

Our approach satisfies the electromagnetic gauge-invariance
constraint. This follows from the on-shell condition for the
generalized potential, for which we will construct a systematic
approximation in the following. The on-shell reaction amplitude
will then be derived in application of
(\ref{Eq:Dispersion_relation}). Owing to the matching scale
$\mu_M$ in (\ref{Eq:Dispersion_relation}) the non-perturbative
coupled-channel calculation and the results from a perturbative
application of the chiral Lagrangian smoothly connect at $s =
\mu_M^2$. This is discussed in detail in \cite{Gasparyan:2010xz}
and \cite{Danilkin:2011fz}. Here we identify $\mu_M$ with the
smallest two-body hadronic threshold value and thereby assume the
applicability of $\chi$PT at  $s = \mu_M^2$.

\begin{table}[t]
\centering \caption{\label{Tab:Left-hand_singularities}The
positions of the closest left-hand
%branch points
branch points of $U_{\textrm{outside}}(s)$ for
$\gamma\gamma\rightarrow PP$ that are determined by the $t$- and
$u$-channel exchange processes. The numbers in the column ``ch =
ab'' correspond to the out-states (a) and in-states (b) of Table
\ref{Tab:Isospin_states_GGtoPP}.}
%\renewcommand{\arraystretch}{1.8}
%\fontsize{8}{2}
\begin{tabular*}{\columnwidth}{@{\extracolsep{\fill}}ccccl@{}}
\hline\noalign{\smallskip}
 $I^G$ & ch. & $\mu_E^2$    & $\Lambda_0^2$                                    & Description \\
\noalign{\smallskip}\hline\noalign{\smallskip}
 $0^+$            & 21  & $m_\pi^2$  &  $-\frac{9\,m_{\pi}^2}{4}\vphantom{\int\limits^M}$                          & $t,u\textrm{-ch}~(2\pi)$ \\[0.5em]
                                    & 31  & $m_K^2$    &  $-\frac{m_\pi^2\,(m_\pi+2\,m_K)^2}{(m_\pi+m_K)^2}$ & $t,u\textrm{-ch}~(\pi K)$ \\[0.7em]
                                    & 41  & $m_\eta^2$ & $0$ & $ t,u\textrm{-ch}~(2\pi)$ \\[0.5em]
%$-\frac{(m_\eta^2-m_\rho^2)^2}{m_\rho^2}$        & $t,u\textrm{-ch}~(\rho)$ \\[0.5em]
\noalign{\smallskip}\hline\noalign{\smallskip}
 $1^-$                    & 21  & $\frac{(m_\pi+m_\eta)^2}{4}\vphantom{\int\limits^M}$ & $\frac{3}{4}\,(m_\eta^2-4\,m_\pi^2)$   & $t,u\textrm{-ch}~(2\pi)$ \\[0.5em]
                                    & 31  & $m_K^2$    &  $-\frac{m_\pi^2\,(m_\pi+2\,m_K)^2}{(m_\pi+m_K)^2}$ & $t,u\textrm{-ch}~(\pi K)$ \\[0.5em]
\noalign{\smallskip}\hline\noalign{\smallskip}
 $2^+$                    & 21  & $m_\pi^2$  &  $-\frac{9\,m_{\pi}^2}{4}\vphantom{\int\limits^M}$ & $t,u\textrm{-ch}~(2\pi)$ \\[0.5em]
\noalign{\smallskip}\hline
\end{tabular*}
\end{table}

Following \cite{Gasparyan:2010xz} the generalized potential
$U_{ab}^J(s)$ can be extrapolated to higher energies in a
controlled manner  by applying conformal mapping techniques. In a
first step the generalized potential is split into two
contributions \footnote{In the following we do not display any
more the angular-momentum superscript $J$ explicitly. Where not
needed we also do not display the channel index $ab$.}
\begin{eqnarray}\label{Eq:U_expansion}
U(s)&=&U_{\textrm{inside}}(s)+U_{\textrm{outside}}(s)  \,,
\end{eqnarray}
where $U_{\textrm{inside}}(s)$ contains the contributions from
close-by left-hand cuts and $U_{\textrm{outside}}(s)$ the
contributions from far-distant left-hand cuts. While the former
can be explicitly calculated from the chiral Lagrangian in a
perturbative application, the latter reflect short distance
physics that need to be parameterized systematically and
efficiently. For the specific example reaction $\gamma \gamma \to
\pi \,\pi$ the separation of 'inside' and 'outside' is implied by
the dashed line in Fig.\ \ref{Fig:Mapping_GGPP}.

The outside potential is expanded in powers of a conformal
variable $\xi(s)$ constructed such as to ensure convergence for
any value of $s$ inside the area bounded by the dashed line of
Fig.\ \ref{Fig:Mapping_GGPP},
\begin{eqnarray}\label{Eq:U_expansion2}
U_{\textrm{outside}}(s)&=&\sum_{k=0}^{n} \,c_k \,\xi^k(s)\quad
\textrm{for}\quad s < \Lambda_s^2 \,,
\end{eqnarray}
where the coefficients $c_k$ are uniquely determined by the first
$k$ derivatives of $U_{\textrm{outside}}(s)$ at the expansion
point $s=\mu^2_E$. We identify this expansion point with the mean
of the initial and the final threshold, assuming that the
chiral expansion is valid there. Following \cite{Danilkin:2011fz}
the function $\xi(s)$ takes the form
\begin{eqnarray} \label{def-conformal}
&&\xi(s)=\frac{a\,(\Lambda^2_s-s)^2-1}{(a-2\,b)(\Lambda^2_s-s)^2+1}\,,\\
&&a = \frac{1}{(\Lambda^2_s-\mu_E^2)^2},\quad b =
\frac{1}{(\Lambda^2_s-\Lambda^2_0)^2}\,,\nonumber
\end{eqnarray}
where the parameter $\Lambda_0$ is determined by the positions of
the closest left-hand branch point of the structures which enter
$U_{\textrm{outside}}$. The quantities $\mu_E$ (expansion point
for outside potential) are collected in Table
\ref{Tab:Left-hand_singularities} for the isospin states of Table
\ref{Tab:Isospin_states_GGtoPP}. Finally, $\Lambda_s$ is the upper
limit of the convergence region, see Fig.\ \ref{Fig:Mapping_GGPP}.

To be specific,  the 'inside' part of the potential receives
contributions from the one-pion (kaon) exchange processes only. We
evaluate the contributions from the cuts starting from
\begin{eqnarray}
&&-\frac{9\,m_\pi^2}{4}< s <0 \qquad \qquad  \qquad \;\;\,{\rm
for} \quad  \gamma\gamma\rightarrow \pi\pi \,,
\nonumber\\
&& -m_\pi^2 \,\frac{(m_\pi+2\,m_K)^2}{(m_\pi+m_K)^2}<s<0 \quad
{\rm for} \quad  \gamma\gamma\rightarrow K\bar K \,, \label{}
\end{eqnarray}
following the procedure outlined in Appendix B of
\cite{Gasparyan:2010xz}. According to Table \ref{Tab:C[SG]} there
are no more cases to be considered. The coefficients $c_k$ in the
outside part of the potential are computed by evaluating the first
$n$ derivatives of the partial-wave amplitudes as determined via
(\ref{Eq:P.w.amplitudes}, \ref{Eq:d-functions}) by the tree-level
result (\ref{Eq:F1F2}).

Via (\ref{Eq:U_expansion}) we obtain an approximated generalized
potential for energies $\Lambda_0^2 < s < \Lambda_s^2$. While the
inside part of the potential is defined for $s >\Lambda^2_s$ also,
the outside part is undefined for $s >\Lambda^2_s$ by
(\ref{Eq:U_expansion2}). We may simply cut off the integral in
(\ref{Eq:Dispersion_relation}) at $\bar s = \Lambda_s^2$. However,
it is advantageous not to do so, since that would induce rapid
variations of the amplitudes close to and below $s = \Lambda_s^2$.
While the precise form of the generalized potential at $s
>\Lambda^2_s$ should not influence the reaction amplitudes in the
target region, where we have a controlled expansion, it is useful
to minimize its residual influence on the target region. This is
the case if the outside potential is smoothly extended for $s >
\Lambda_s^2$ by a constant (see
\cite{Gasparyan:2010xz,Danilkin:2010xd,Danilkin:2011fz,Gasparyan:2011yw}).
We note that due to the particular form of the conformal map
(\ref{def-conformal}), the generalized potential and its
derivative are continuous at $s = \Lambda_s^2$.

Following \cite{Danilkin:2011fz,Danilkin:2012ap}, we choose in the
following $n=3$ in (\ref{Eq:U_expansion2}) and
$\Lambda_s=1.6\,$GeV. We have varied $\Lambda_s$ in the range 1.4
GeV to 1.8 GeV. The impact of this variation on the results is
small \cite{Danilkin:2012ap}.

\begin{table}
\centering \caption{\label{Tab:C[SG]}The coefficients $C_{SG}$,
$C_{0}^{(x)} $ and $C_{1}^{(x)} = 2 \, C_{h_P}^{(x)}$ of the
invariant amplitudes (\ref{Eq:F1F2}) with respect to the
coupled-channel states $I^G$ of Table
\ref{Tab:Isospin_states_GGtoPP}. The numbers in the column ``ch =
ab'' correspond to the out-states (a) and in-states (b) of Table
\ref{Tab:Isospin_states_GGtoPP}.}
\begin{tabular*} {\columnwidth}{@{\extracolsep{\fill}}ccccc@{}}
\hline\noalign{\smallskip}
%\multicolumn{2}{c}{$0^+$}
%$I^G$ & \multicolumn{2}{c}{$0^+$} & $1^-$ & $2^+$ \\
%
$I^G$ & $0^+$ & $0^+$ & $1^-$ & $2^+$ \\
\noalign{\smallskip}\hline\noalign{\smallskip}
ch.   & 21    &31 &31&21\\
\noalign{\smallskip}\hline\noalign{\smallskip}
$C_{SG}$ & $-\frac{8}{\sqrt{3}}$ & $-4$ & $-4$ &
$4\sqrt{\frac{2}{3}}$\\
\noalign{\smallskip}\hline\noalign{\smallskip}
$C^{(\pi)}_{0}$ &$-\frac{8}{\sqrt{3}}$ & 0& 0& $4\sqrt{\frac{2}{3}}$\\
\noalign{\smallskip}\hline\noalign{\smallskip}
$C^{(K)}_{0}$ & 0& $-4$ & $-4$ & 0\\
\noalign{\smallskip}\hline\noalign{\smallskip}
$C_1^{(\rho)}$& $\frac{16}{\sqrt{3}}$& 4& 4& $-8\sqrt{\frac{2}{3}}$\\
\noalign{\smallskip}\hline\noalign{\smallskip}
$C_1^{(\omega)}$& 0& $\frac{4}{3}$& $\frac{4}{3}$& 0\\
\noalign{\smallskip}\hline\noalign{\smallskip}
$C_1^{(\phi)}$& 0& $\frac{8}{3}$& $\frac{8}{3}$& 0\\
\noalign{\smallskip}\hline
\end{tabular*}
\end{table}

\begin{table*}[htbp]
\centering \caption{\label{Tab:C[bAhA]}The coefficients
$C_{b_A\,h_A}^{(x,y,z)} = C_{b_A\,h_O}^{(x,y,z)}  $. See the
caption of Table \ref{Tab:C[SG]} for more details.}
\begin{tabular*}{\textwidth}{@{\extracolsep{\fill}}cccccccc@{}}
\hline\noalign{\smallskip}
\multicolumn{8}{c}{$C_{b_A\,h_A}^{(x,y,z)}$}\\
\noalign{\smallskip}\hline\noalign{\smallskip}
 $I^G$ & ch. &
$(\rho,\rho,\rho)$&$(\rho,\rho,\omega)$ & $(\rho,\rho,K^*)$ &
$\begin{array}{l}(\rho,\omega,\rho)\\(\omega,\rho,\omega)\end{array}$
&
$\begin{array}{l}(\rho,\omega,K^*)\\(\omega,\rho,K^*)\end{array}$&
$\begin{array}{l}(\rho,\phi,K^*)\\(\phi,\rho,K^*)\end{array}$\\
\noalign{\smallskip}\hline\noalign{\smallskip}
$0^+$        & 21 & 0                      & $\frac{32\,m_\pi^2}{\sqrt{3}}$ & 0          & 0                                        & 0                    & 0                             \\
                 & 31 & 0                      & 0                              & $32\,m_K^2$& 0                                        & 0                    & 0                             \\
                 & 41 & $\frac{32\,m_\pi^2}{3}$& 0                              & 0          & 0                                        & 0                    & 0                             \\
\noalign{\smallskip}\hline\noalign{\smallskip}
 $1^-$ & 21 & 0                      & 0                              & 0          & $\frac{32}{3}\sqrt{\frac{2}{3}}\,m_\pi^2$& 0                    & 0                             \\
                 & 31 & 0                      & 0                              & 0          & 0                                        & $\frac{16\,m_K^2}{3}$& $-\frac{32\,m_K^2}{3}$        \\
\noalign{\smallskip}\hline\noalign{\smallskip}
 $2^+$  & 21 & 0                      & $32\sqrt{\frac{2}{3}}\,m_\pi^2$& 0          & 0                                        & 0                    & 0                             \\
\noalign{\smallskip}\hline\noalign{\smallskip}
  & &
$(\omega,\omega,\rho)$&$(\omega,\omega,\omega)$ &
$(\omega,\omega,K^*)$ &
$\begin{array}{l}(\omega,\phi,K^*)\\(\phi,\omega,K^*)\end{array}$&
$(\phi,\phi,K^*)$ & $(\phi,\phi,\phi)$ \\
\noalign{\smallskip}\hline\noalign{\smallskip}
$0^+$        & 21 & $\frac{32\,m_\pi^2}{3\sqrt{3}}$& 0                       & 0                    & 0                     & 0                     & 0                             \\
                 & 31 & 0                              & 0                       & $\frac{16\,m_K^2}{9}$& $-\frac{32\,m_K^2}{9}$& $\frac{64\,m_K^2}{9}$& 0                             \\
                 & 41 & 0                              & $\frac{32\,m_\pi^2}{27}$& 0                    & 0                     & 0                     & $\frac{32}{27}\,(16\,m_K^2-8\,m_\pi^2)$\\
%\hline $1^-$ & 13 & 0                              & 0                       & 0                    & 0                     & 0                     & 0                             \\
%                 & 23 & 0                              & 0                       & 0                    & 0                     & 0                     & 0                             \\
%\hline$2^+$  & 12 & 0                              & 0                       & 0                    & 0                     & 0                     & 0                             \\
\noalign{\smallskip}\hline
\end{tabular*}
\end{table*}

A crucial observation behind our summation scheme is the fact that
the computation of the partial-wave scattering amplitude from
(\ref{Eq:Dispersion_relation}) at energies larger than threshold
also requires only the knowledge of the generalized potential at
energies larger than threshold. More generally, depending on where
we want to compute the partial-wave scattering amplitudes, it
suffices to construct a controlled approximation of the
generalized potential in a specific region of the complex plane
only. This is always achieved with (\ref{Eq:U_expansion},
\ref{Eq:U_expansion2}) and the desired solution of
(\ref{Eq:Dispersion_relation}) can be found by the $N/D$ ansatz
\cite{Chew:1960iv}
\begin{eqnarray}\label{Eq:NoverD}
T_{ab}(s)=\sum_{c}\,D^{-1}_{ac}(s)\,N_{cb}(s)\,,
\end{eqnarray}
where $D_{ab}(s)$ contains only the right-hand s-channel unitarity
cuts,
\begin{eqnarray}\label{Eq:D_function}
&&D_{ab}(s)=\delta_{ab}-\sum_{c}\int_{\mu_{\rm
thr}^2}^{\infty}\frac{d\bar{s}}{\pi}\frac{s-\mu_M^2}{\bar{s}-\mu_M^2}
\frac{N_{ac}(\bar{s})\rho_{cb}(\bar{s})}{\bar{s}-s-i\epsilon}\,,
\end{eqnarray}
and the matrix $N_{ab}(s)$ contains left-hand cuts only,
\begin{eqnarray}\label{Eq:N_function}
&& N_{ab}(s)=U_{ab}(s)\\
&&+\sum_{c,d}\int_{\mu_{\rm thr}^2}^\infty
\frac{d\bar{s}}{\pi}\,\frac{s-\mu^2_M}{\bar{s}-\mu^2_M}\,N_{ac}(\bar{s})\,\rho_{cd}(\bar{s})\,\frac{U_{db}(\bar{s})-U_{db}(s)}{\bar{s}-s}\,.
\nonumber
\end{eqnarray}
The system (\ref{Eq:N_function}) with the input
(\ref{Eq:U_expansion}), \eqref{Eq:U_expansion2} can be solved
numerically by the method of matrix inversion.

It is instructive to compare our approach with dispersive studies based on the Roy-Steiner equations
\cite{Roy:1971tc,Ananthanarayan:2000ht,Buettiker:2003pp,GarciaMartin:2011jx,Moussallam:2011zg,Hoferichter:2011wk}.
While the latter analyses aim at constraining the low-energy scattering amplitude by using
experimental input at high energies, we perform an analytic continuation of the subthreshold amplitudes to higher energies,
where resonances may play an important role. The subthreshold amplitudes are computed in a conformal expansion based on
the chiral Lagrangian. The analytic extrapolation is implied by the solution of the non-linear integral
equation (\ref{Eq:Dispersion_relation}). Due to the analytic continuation we can compute the partial-wave amplitudes
only in a specific domain in the complex plane (see Fig. 2). That implies that the consequences of crossing symmetry can
not be verified directly everywhere in our approach. Only in a small subthreshold window crossing symmetry is directly
testable. However, by construction there our amplitudes are well approximated by perturbative expressions that respect the
constraints of crossing symmetry manifestly \cite{Lutz:2003fm}. To this extent crossing symmetry is satisfied in
our approach in an approximate manner, the accuracy of which is expected to increase more and more as higher
order effects are considered in the computation. This is contrasted by the Roy-Steiner equations that implement crossing symmetry
exactly. The benefit of our approach is, that the treatment of many coupled channels is quite feasible. Applying Roy-Steiner
equations for many channels is quite complicated and so far has not been achieved in the literature.

\begin{table*}[t]
\centering \caption{\label{Tab:C[bA,hA]}The coefficients
$C_{b_A}^{(x,y,z)}$ and $C_{h_A}^{(x,y,z)}  = C_{h_O}^{(x,y,z)} $.
See the caption of Table \ref{Tab:C[SG]} for more details.}
%\small
\begin{tabular*}{\textwidth}{@{\extracolsep{\fill}}cccccccc@{}}
\hline\noalign{\smallskip}
\multicolumn{8}{c}{$C_{b_A}^{(x,y,z)}$}\\
\noalign{\smallskip}\hline\noalign{\smallskip}
 $I^G$ & ch. & $(\rho,\rho,\rho)$&$(\rho,\rho,\omega)$
& $(\rho,\rho,K^*)$ &
$\begin{array}{l}(\rho,\omega,\rho)\\(\omega,\rho,\omega)\end{array}$
&
$\begin{array}{l}(\rho,\omega,K^*)\\(\omega,\rho,K^*)\end{array}$
&
$\begin{array}{l}(\rho,\phi,K^*)\\(\phi,\rho,K^*)\end{array}$\\
\noalign{\smallskip}\hline\noalign{\smallskip}
$0^+$        & 21 & 0                      & $\frac{64\,m_\pi^4}{\sqrt{3}}$ & 0          & 0                                        & 0                    & 0                             \\
                 & 31 & 0                      & 0                              & $32\,m_K^4$& 0                                        & 0                    & 0                             \\
                 & 41 & $\frac{64\,m_\pi^4}{3}$& 0                              & 0          & 0                                        & 0                    & 0                             \\
\noalign{\smallskip}\hline\noalign{\smallskip}
 $1^-$ & 21 & 0                      & 0                              & 0          & $\frac{64}{3}\sqrt{\frac{2}{3}}\,m_\pi^4$& 0                    & 0                             \\
                 & 31 & 0                      & 0                              & 0          & 0                                        & $\frac{32\,m_K^4}{3}$& $-\frac{64\,m_K^4}{3}$        \\
\noalign{\smallskip}\hline\noalign{\smallskip}
$2^+$  & 21 & 0                      & $64\sqrt{\frac{2}{3}}\,m_\pi^4$& 0          & 0                                        & 0                    & 0                             \\
\noalign{\smallskip}\hline\noalign{\smallskip} & &
$(\omega,\omega,\rho)$ & $(\omega,\omega,\omega)$ &
$(\omega,\omega,K^*)$ &
$\begin{array}{l}(\omega,\phi,K^*)\\(\phi,\omega,K^*)\end{array}$&
$(\phi,\phi,K^*)$ & $(\phi,\phi,\phi)$ \\
\noalign{\smallskip}\hline\noalign{\smallskip}
$0^+$        & 21 & $\frac{64\,m_\pi^4}{3\sqrt{3}}$& 0                       & 0                    & 0                     & 0                     & 0                             \\
                 & 31 & 0                              & 0                       & $\frac{32\,m_K^4}{9}$& $-\frac{64\,m_K^4}{9}$& $\frac{128\,m_K^4}{9}$& 0                             \\
                 & 41 & 0                              & $\frac{64\,m_\pi^4}{27}$& 0                    & 0                     & 0                     & $\begin{array}{l}
                 \frac{64}{27}\,(32\,m_K^4-\\ 32\,m_\pi^2\,m_K^2+8\,m_\pi^4) \end{array}$\\
%\hline $(1^-,0)$ & 13 & 0                              & 0                       & 0                    & 0                     & 0                     & 0                             \\
%                 & 23 & 0                              & 0                       & 0                    & 0                     & 0                     & 0                             \\
%\hline$(2^+,0)$  & 12 & 0                              & 0                       & 0                    & 0                     & 0                     & 0                             \\
\noalign{\smallskip}\hline\noalign{\smallskip}
\multicolumn{8}{c}{$C_{h_A}^{(x,y,z)}$}\\
\noalign{\smallskip}\hline\noalign{\smallskip}
 $I^G$ & ch. &
$(\rho,\rho,\rho)$&$(\rho,\rho,\omega)$ & $(\rho,\rho,K^*)$ &
$\begin{array}{l}(\rho,\omega,\rho)\\(\omega,\rho,\omega)\end{array}$&
$\begin{array}{l}(\rho,\omega,K^*)\\(\omega,\rho,K^*)\end{array}$&
$\begin{array}{l}(\rho,\phi,K^*)\\(\phi,\rho,K^*)\end{array}$\\
\noalign{\smallskip}\hline\noalign{\smallskip}
$0^+$        & 21 & 0              & $\frac{16}{\sqrt{3}}$  & 0  & 0                               & 0            & 0                    \\
                 & 31 & 0              & 0                      & $8$& 0                               & 0            & 0                    \\
                 & 41 & $\frac{16}{3}$ & 0                      & 0  & 0                               & 0            & 0                    \\
\noalign{\smallskip}\hline\noalign{\smallskip}
 $1^-$ & 21 & 0              & 0                      & 0  & $\frac{16}{3}\sqrt{\frac{2}{3}}$& 0            & 0                    \\
                 & 31 & 0              & 0                      & 0  & 0                               & $\frac{8}{3}$& $-\frac{16}{3}$      \\
\noalign{\smallskip}\hline\noalign{\smallskip}
 $2^+$  & 21 & 0              & $16 \sqrt{\frac{2}{3}}$& 0  & 0                               & 0            & 0                    \\
\noalign{\smallskip}\hline\noalign{\smallskip}
 & &
$(\omega,\omega,\rho)$&$(\omega,\omega,\omega)$ &
$(\omega,\omega,K^*)$ &
$\begin{array}{l}(\omega,\phi,K^*)\\(\phi,\omega,K^*)\end{array}$&
$(\phi,\phi,K^*)$ & $(\phi,\phi,\phi)$ \\
\noalign{\smallskip}\hline\noalign{\smallskip}
$0^+$        & 21 & $\frac{16}{3 \sqrt{3}}$& 0              & 0            & 0              & 0             & 0                             \\
                 & 31 & 0                      & 0              & $\frac{8}{9}$& $-\frac{16}{9}$& $\frac{32}{9}$& 0                             \\
                 & 41 & 0                      & $\frac{16}{27}$& 0            & 0              & 0             & $\frac{128}{27}$\\
%\hline $(1^-,0)$ & 13 & 0                      & 0              & 0            & 0              & 0             & 0                             \\
%                 & 23 & 0                      & 0              & 0            & 0              & 0             & 0                             \\
%\hline$(2^+,0)$  & 12 & 0                      & 0              & 0            & 0              & 0             & 0                             \\
\noalign{\smallskip}\hline
\end{tabular*}
\end{table*}

\begin{table*}[htbp]
\renewcommand{\arraystretch}{1.2}
\centering \caption{\label{Tab:C[g1,g2,g3,g5]}The coefficients
$C_{g_1}^{(x,y)}$, $C_{g_2}^{(x,y)}$, $C_{g_3}^{(x,y)}$,
$C_{g_5}^{(x,y)}$ and $C_{b_D}^{(x,y)}$. See the caption of Table
\ref{Tab:C[SG]} for more details.}
\begin{tabular*}{\textwidth}{@{\extracolsep{\fill}}cccccccc@{}}
\hline\noalign{\smallskip}
$I^G$ & ch. & $(\rho,\rho)$ &
$\begin{array}{l}(\rho,\omega)\\(\omega,\rho)\end{array}$ &
$\begin{array}{l}(\rho,\phi)\\(\phi,\rho)\end{array}$ &
$(\omega,\omega)$ &
$\begin{array}{l}(\omega,\phi)\\(\phi,\omega)\end{array}$ & $(\phi,\phi)$ \\
\noalign{\smallskip}\hline\noalign{\smallskip}
\multicolumn{8}{c}{$C_{g_1}^{(x,y)}$}\\
\noalign{\smallskip}\hline\noalign{\smallskip}
$0^+$            &   21&  $\frac{16}{\sqrt{3}}$ & 0                                &0               &$\frac{16}{3 \sqrt{3}}$& 0               & 0                \\
                 &   31&  $8$                   & 0                                &0               &$\frac{8}{9}$          & $-\frac{16}{9}$ & $\frac{32}{9}$   \\
                 &   41&  $16/3$                & 0                                &0               &$\frac{16}{27}$        & 0               & $\frac{128}{27}$ \\
\noalign{\smallskip}\hline\noalign{\smallskip}
$1^-$            &   21&  0                     & $\frac{16}{3}\sqrt{\frac{2}{3}}$ &0               &0                      & 0               & 0                \\
                 &   31&  0                     & $\frac{8}{3}$                    &$-\frac{16}{3}$ &0                      & 0               & 0                \\
\noalign{\smallskip}\hline\noalign{\smallskip}
$2^+$            &   21& $16 \sqrt{\frac{2}{3}}$& 0                                &   0            &0                      & 0               & 0                \\
\noalign{\smallskip}\hline\noalign{\smallskip}
\multicolumn{8}{c}{$C_{g_2}^{(x,y)}$}\\
\noalign{\smallskip}\hline\noalign{\smallskip}
 $0^+$
                 &   21&  $\frac{32}{\sqrt{3}}$  & 0                                &0               &0                      & 0               & 0                \\
                 &   31&  $8$                    & 0                                &0               &$\frac{8}{9}$          & $\frac{16}{9}$  & $\frac{32}{9}$   \\
\noalign{\smallskip}\hline\noalign{\smallskip}
$1^-$            &   21&  0                      & $\frac{8}{3}$                    &$\frac{16}{3}$  &0                      & 0               & 0                \\
\noalign{\smallskip}\hline\noalign{\smallskip}
$2^+$            &   21& $-16 \sqrt{\frac{2}{3}}$& 0                                &   0            &0                      & 0               & 0                \\
\noalign{\smallskip}\hline\noalign{\smallskip}
\multicolumn{8}{c}{$C_{g_3}^{(x,y)}$}\\
\noalign{\smallskip}\hline\noalign{\smallskip}
 $0^+$
                 &   21&  $4 \sqrt{3}$ & 0                              & 0  & $\frac{4}{3 \sqrt{3}}$  & 0 &0\\
                 &   31&  $4$          & 0                              & 0  & $\frac{4}{9}$           & 0 &$\frac{16}{9}$\\
                 &   41&  $\frac{4}{3}$& 0                              & 0  & $\frac{4}{27}$          & 0 &$\frac{32}{27}$\\
\noalign{\smallskip}\hline\noalign{\smallskip}
 $1^-$           &   21&  0            & $\frac{4}{3}\sqrt{\frac{2}{3}}$& 0  &0                        & 0 &0   \\
                 &   31&  0            & $\frac{4}{3}$                  & 0  &0                        & 0 &0   \\
\noalign{\smallskip}\hline\noalign{\smallskip}
\multicolumn{8}{c}{$C_{g_5}^{(x,y)}$}\\
\noalign{\smallskip}\hline\noalign{\smallskip}
 $0^+$
                 &   21&  $-\frac{8}{\sqrt{3}}$ & 0                                &0               &0                      & 0               & 0                \\
                 &   31&  -2                    & 0                                &0               &$-\frac{2}{9}$         & $-\frac{4}{9}$  & $-\frac{8}{9}$    \\
\noalign{\smallskip}\hline\noalign{\smallskip}
 $1^-$           &   31&  0                      & $-\frac{2}{3}$                   &$-\frac{4}{3}$  &0                      & 0               & 0                \\
\noalign{\smallskip}\hline\noalign{\smallskip}
 $2^+$           &   21& $4\sqrt{\frac{2}{3}}$  & 0                                &   0            &0                      & 0               & 0                \\
\noalign{\smallskip}\hline\noalign{\smallskip}

\multicolumn{8}{c}{$C_{b_D}^{(x,y)}$}\\
\noalign{\smallskip}\hline\noalign{\smallskip}
 $0^+$
                 &   21&  $32\sqrt{3}\,m_\pi^2$  & 0                                         & 0& $\frac{32}{3 \sqrt{3}}\,m_\pi^2$  & 0& 0                \\
                 &   31&  $32\,m_K^2$            & 0                                         & 0& $\frac{32}{9}\,m_K^2$             & 0& $\frac{128}{9}\,m_K^2$   \\
                 &   41&  $\frac{32}{3}\,m_\pi^2$& 0                                         & 0& $\frac{32}{27}\,m_\pi^2$          & 0& $\frac{32}{27} \left(16\,m_K^2-8\,m_\pi^2\right)$\\
\noalign{\smallskip}\hline\noalign{\smallskip}
 $1^-$           &   21&  0                      & $\frac{32}{3}\sqrt{\frac{2}{3}}\,m_\pi^2$ & 0&0                                  & 0& 0                \\
                 &   31&  0                      & $\frac{32}{3}\,m_K^2$                     & 0&0                                  & 0& 0                \\
\noalign{\smallskip}\hline
\end{tabular*}
\end{table*}

\section{Numerical results}
\label{sec:numres}

In this section we present our results for the cross
sections\footnote{Usually the experimental results are limited to
a range of $|x|\leq Z$ with $x= \cos\theta$. In this case the
cross section is given by
$\sigma=2\int\limits_{0}^Z\,\frac{d\sigma}{dx}\,dx$.} of the
reactions $\gamma\gamma\rightarrow \pi^0\pi^0$, $\pi^+\pi^-$,
$K^0\bar{K}^0$, $K^+K^-$, $\eta\,\eta$ and $\pi^0\eta$, evaluated
with the $J = 0,2$ partial-waves amplitudes. We have checked that
the contributions from the higher partial waves are negligible in
the energy range $\sqrt{s}< 1.2$ GeV.

\begin{figure*}[t]
\begin{center}
\vskip-0.1cm \hskip-0.5cm
\parbox{17.0cm}{\includegraphics[keepaspectratio,width=\textwidth]{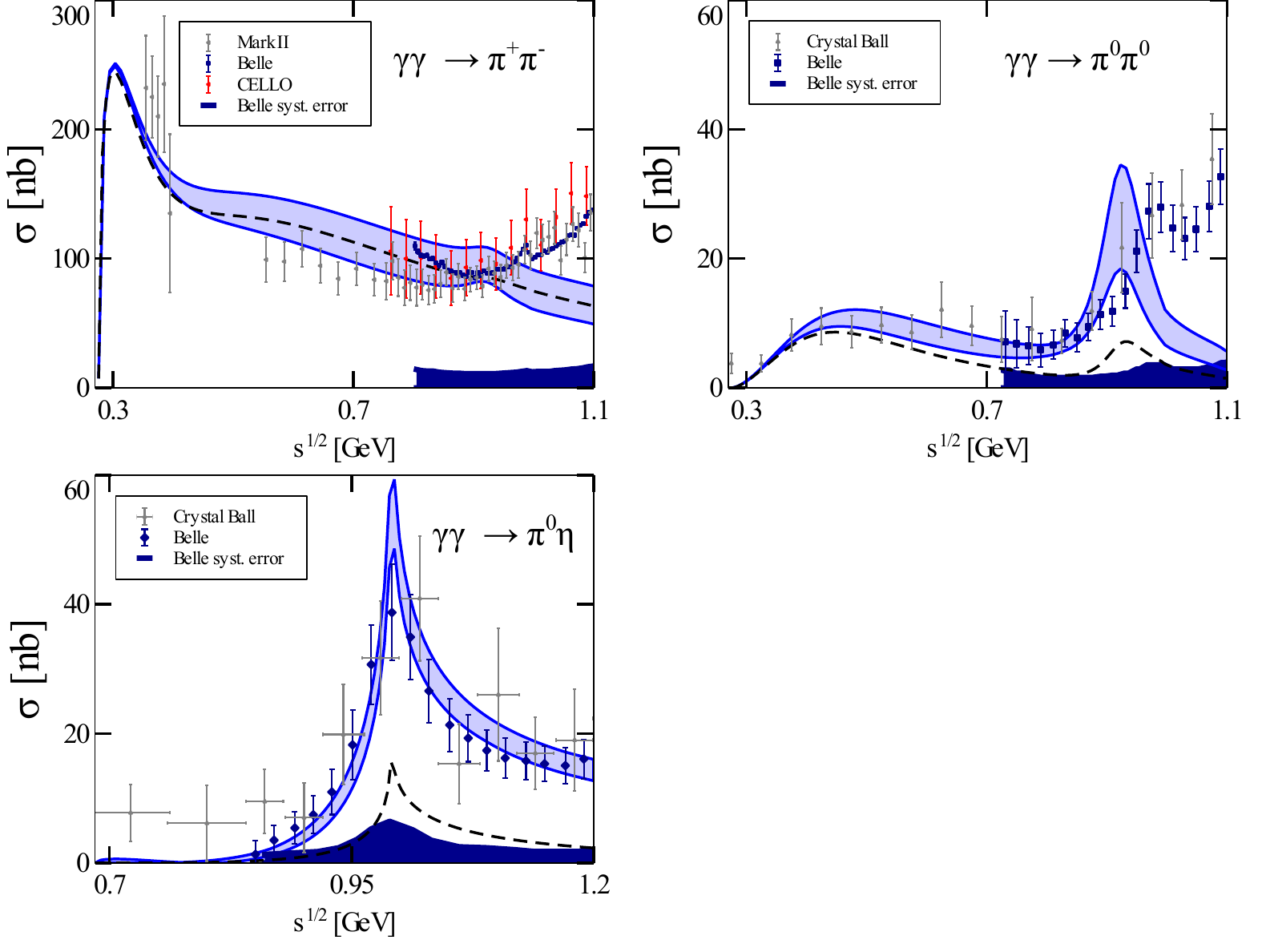}%
}
\parbox{13.5cm}{\vspace{-9cm}
\hspace*{7.3cm}
\parbox{7.5cm}{\caption{\label{Fig:Cross_sections_GGtoPiPi_different_g3g5hO}Total cross
sections for $\gamma\gamma\rightarrow \pi^+\pi^-$ with
$|\cos\theta|<0.6$ (top left), $\gamma\gamma\rightarrow
\pi^0\pi^0$ with $|\cos\theta|<0.8$ (top right) and
$\gamma\gamma\rightarrow \pi^0\eta$ with $|\cos\theta|<0.9$
(bottom). A variation of parameters $g_3,\,g_5,\,h_O\in[-5,5]$
using (\ref{Eq:Correlation of parameters}) is reflected by the
various bands. Setting $g_i = 0 = h_O$ yields the dashed lines.
The data are taken from
\cite{Marsiske:1990hx,Uehara:2009cka,Boyer:1990vu,Behrend:1992hy,Mori:2007bu,Antreasyan:1985wx,Uehara:2009cf}.}}}
\end{center}
\end{figure*}

We use the set of parameters given in (\ref{Eq:Set_of_parameters})
for all the numerical results. However, the remaining five
parameters $g_1$, $g_2$, $g_3$, $g_5$ and $h_O$ have to be
determined. Our strategy is to use the empirical data on the
reactions $\gamma\gamma\rightarrow\pi^0\pi^0$, $\pi^+\pi^-$ and
$\pi^0\eta$ and in addition the differential and integrated data
for the decay $\eta\rightarrow \pi^0 \gamma \gamma$. The results
for the cross sections of the reactions $\gamma\gamma\rightarrow
K^0\bar{K}^0$, $K^+K^-$ and $\eta\,\eta$ are then pure
predictions.

On account of crossing symmetry, the decay amplitude for
$\eta\rightarrow \pi^0 \gamma \gamma$ can be easily obtained from
$\gamma \gamma \rightarrow \pi^0 \eta$ by considering $\pi^0$ and
photons as outgoing particles. To get the decay amplitude it is
enough to replace
\begin{eqnarray}
  s &=& (k_1+k_2)^2  \rightarrow (\bar{k}_1+\bar{k}_2)^2= M_{\gamma\gamma}^2\,,\nonumber\\
  t &=& (p-k_1)^2  \rightarrow (p_\pi+\bar{k}_1)^2= M_{\gamma_1\pi}^2\,,\\
  u &=& (p-k_2)^2  \rightarrow (p_\pi+\bar{k}_2)^2= M_{\gamma_2\pi}^2\,,\nonumber
\end{eqnarray}
in the invariant amplitudes (\ref{Eq:F1F2}). The differential
decay rate is given by \cite{Nakamura:2010zzi}
\begin{equation}\label{Eq:Eta_decay}
  d\Gamma=\frac{1}{(2\pi)^3}\,\frac{1}{32\,m_\eta^3}\,\sum_{pol}|T_{\eta\rightarrow\pi^0\gamma\gamma}|^2\,
  dM_{\gamma\gamma}^2\,dM_{\gamma_2\pi}^2  \,.
\end{equation}
For the integrated partial decay width one has to include the
degeneracy factor of 1/2 to account for the fact that one has two
indistinguishable photons in the final state. To obtain the {\em
decay} amplitude we use directly the tree-level result
\eqref{Eq:F1F2}. Since we are here in the low-energy decay region,
we assume that coupled-channel effects are less important. For the
{\em reaction} amplitudes we use, of course, the full rescattering
formalism outlined in the previous section.

In a first step we use the reaction data of
$\gamma\gamma\rightarrow\pi^0\pi^0$, $\pi^+\pi^-$ and $\pi^0\eta$
to correlate the five free parameters. Having matched the data
with the coupled-channel calculations leads to the following
relations:
\begin{eqnarray}\label{Eq:Correlation of parameters}
  g_1 &=& 0.900 -0.200 \, g_3 + 0.038 \, h_O^2 + 0.128 \, h_O \,, \nonumber\\
  g_2 &=& -1.50 - 0.27 \, g_3 + 0.25 \, g_5 \,.
\end{eqnarray}
This leaves us with three free parameters. If they are varied
within the range $g_3,\,g_5,\,h_O\in[-5,5]$ one obtains the cross
sections depicted in Fig.\
\ref{Fig:Cross_sections_GGtoPiPi_different_g3g5hO}. A detailed
discussion of the cross sections will be given below. To get a
feeling for the influence of the five parameters we also provide
the cross sections for the case where all these five parameters
are put to zero, see the dashed lines in Fig.\
\ref{Fig:Cross_sections_GGtoPiPi_different_g3g5hO}. Obviously, one
would significantly underestimate the data in both neutral
channels $\pi^0 \pi^0$ and $\pi^0 \eta$ without the parameters
$g_i$ and $h_O$. Note, however, that the qualitative structure
does not depend so much on these parameters.

\begin{figure}[h]
\center{\includegraphics[keepaspectratio,width=0.45\textwidth]{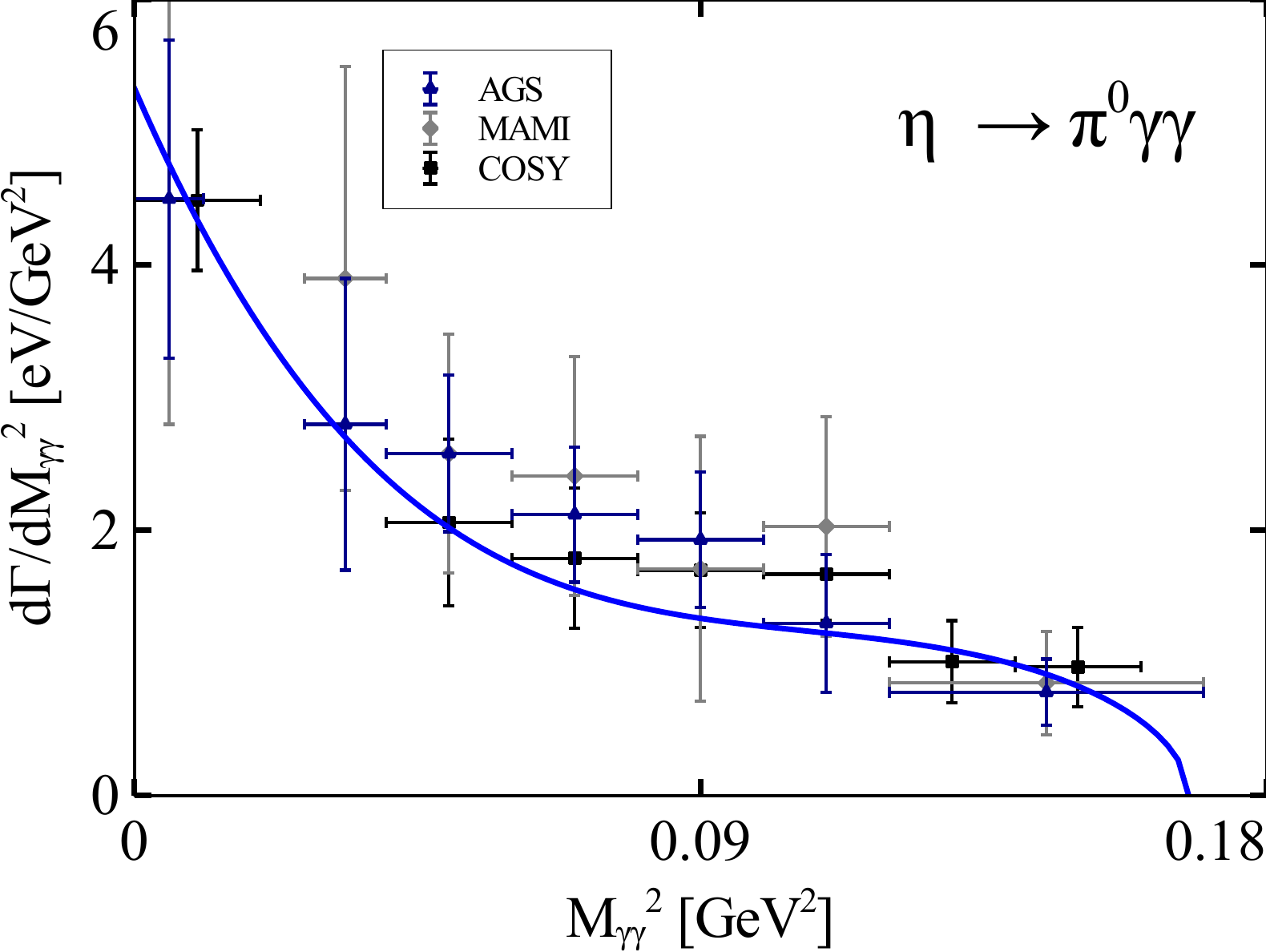}%
}  %
\caption{\label{Fig:Eta_decay}The single-differential
invariant-mass distribution of the decay $\eta\rightarrow \pi^0
\gamma \gamma$. Parameters are chosen according to
\eqref{Eq:Correlation of parameters}, \eqref{Eq:NEW_parameters}.
Note that the parameters $g_2$ and $g_5$ do not contribute to this
decay. The data are taken from
\cite{Prakhov:1900zz,Prakhov:2008zz,Lalwani-PhD}.}
\end{figure}

We continue with a determination of the remaining parameters using
the existing data on $\eta\rightarrow \pi^0 \gamma \gamma$ decay.
The present experimental status for $\eta\rightarrow \pi^0 \gamma
\gamma$ decay is the following: The Particle Data Group
\cite{Nakamura:2010zzi} gives the branching ratio
$\Gamma_{\eta\rightarrow \pi^0 \gamma \gamma}/\Gamma_\eta=(2.7 \pm
0.5) \cdot 10^{-4}$ and the full width $\Gamma_\eta= (1.30\pm
0.07)\,$keV. This results in a partial decay width of
$\Gamma_{\eta\rightarrow \pi^0 \gamma \gamma} \approx (0.35 \pm
0.09)\,$eV. Theoretical studies have been performed in
\cite{Ametller:1991dp,Oset:2002sh,Oset:2008hp}.

For the decay $\eta\rightarrow \pi^0\gamma\gamma$ three of the yet
undetermined parameters contribute, namely $g_1$, $g_3$ and $h_O$.
Using the relation (\ref{Eq:Correlation of parameters}) for $g_1$,
we adjust $g_3$ and $h_O$ to the partial decay width and to the
two-photon invariant-mass distribution depicted in Fig.\
\ref{Fig:Eta_decay}. In this way we find
\begin{eqnarray}
\label{Eq:NEW_parameters} g_3=-4.88 \,, \qquad h_O= 3.27
\end{eqnarray}
which implies $g_1=2.70$ and $g_2=-0.18+0.25\,g_5$. The fit yields
\begin{equation}\label{Eq:Integrated_width}
    \Gamma_{\eta\rightarrow \pi^0 \gamma \gamma}=0.310 \,\textrm{eV}
\end{equation}
for the integrated partial decay width, in good agreement with the
experimental value.

\begin{figure*}[t]
  \centering
  \includegraphics[keepaspectratio,width=\textwidth]{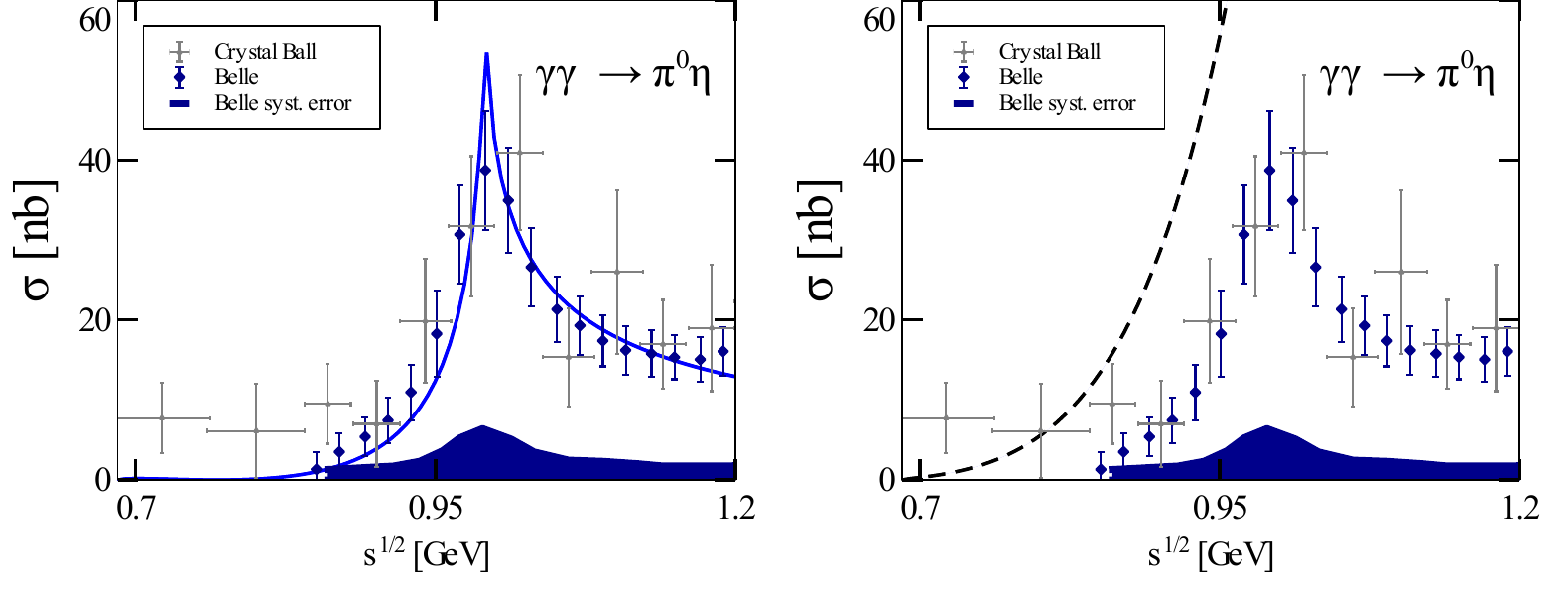}
  \caption{Cross section for the reaction $\gamma\gamma\rightarrow \pi^0\eta$ using \eqref{Eq:Correlation of parameters},
    \eqref{Eq:NEW_parameters} together with $g_5\in[-5,5]$. The full  result is shown on the left and the tree-level result
    on the right. The tree-level result does not depend on $g_5$. Also the dependence of the full result on $g_5$ is very weak.
    It is caused by the cross-channel effect $\gamma\gamma\rightarrow K \bar K \to \pi^0\eta$.
    See the figure caption of Fig.\ \ref{Fig:Cross_sections_GGtoPiPi_different_g3g5hO} for more details.}
  \label{fig:etapi-full-tree}
\end{figure*}

We have determined four of our five free parameters. In the
following we will show results where the remaining free parameter
$g_5$ is varied in the range $g_5\in[-5,5]$. Note that the
achieved determination of the parameters is also crucial for
future investigations. Originally all these parameters concern
interactions between two vector mesons and an odd ($h_O$) or even
($g_i$, $i=1,2,3,5$) number of Goldstone bosons. In the future it
is planned to explore also the importance of vector-meson channels
for the coupled-channel problems (cf.\ the corresponding
discussion in the introduction and \cite{Lutz:2011xc}). There the
coupling constants $h_O$ and $g_i$ enter directly and mediate,
e.g., the transition from two vector to two pseudoscalar mesons.
For the following reason these coupling constants are also
important for our case at hand, in spite of the fact that we do
not consider the vector-meson channels: The neutral vector mesons
couple directly to photons; see also Fig.\
\ref{Fig:Tree_level_diagrams}. Therefore, the coupling constants
$h_O$ and $g_i$ enter also the transition amplitudes from two
photons to two pseudoscalars. In turn, data on such interactions
between hadrons and electromagnetism can be used to constrain
purely hadronic coupling constants. This resembles our
determination of $h_A$ from the decay $\omega \to \gamma \pi^0$ in
\cite{Lutz:2008km}. Note, however, that in our formalism this line
of reasoning does not lead automatically to strict vector-meson
dominance, but rather to an improved version thereof
\cite{Terschluesen:2010ik,Terschlusen:2012xw}.

We now turn to a detailed discussion of the various two-meson
channels populated by photon fusion. The first highlight is the
$\pi^0 \eta$ channel depicted in Fig.\ \ref{fig:etapi-full-tree}.
Here our formalism shows a dynamically generated scalar-isovector
resonance which is in full quantitative agreement with the
experimental data; see also
\cite{Oller:1997ti,Oller:1997yg,Doring:2011vk} where similar
findings have been reported. In our approach we find that this
$a_0(980)$ resonance coincides with the kaon-antikaon threshold and
emerges from rescattering and coupled-channel effects between
$\pi^0 \eta$ and $K \bar K$. We contrast our full coupled-channel
result with a pure tree-level calculation based on
\eqref{Eq:F1F2}. The latter is also depicted in Fig.\
\ref{fig:etapi-full-tree} and, of course, does not show a
resonance shape, in obvious disagreement with the experimental
data. We stress again that according to the hadrogenesis
conjecture
\cite{Lutz:2001dr,Lutz:2003fm,Lutz:2004dz,Lutz:2005ip,Lutz:2007sk,Lutz:2008km,Terschlusen:2012xw}
the low-lying scalar resonances are supposed to be generated
dynamically. An incarnation of this proposition is seen in Fig.\
\ref{fig:etapi-full-tree}. We recall from our previous discussion
about Fig.\ \ref{Fig:Cross_sections_GGtoPiPi_different_g3g5hO}
that the location of the resonance does not depend on the choice
of the coupling constants $g_i$ and $h_O$. Only the height of the
curve is sensitive to these parameters. This provides confidence
in the robustness of our interpretation of the lowest-lying
scalar-isovector resonance.

\begin{figure*}
  \centering
  \includegraphics[keepaspectratio,width=\textwidth]{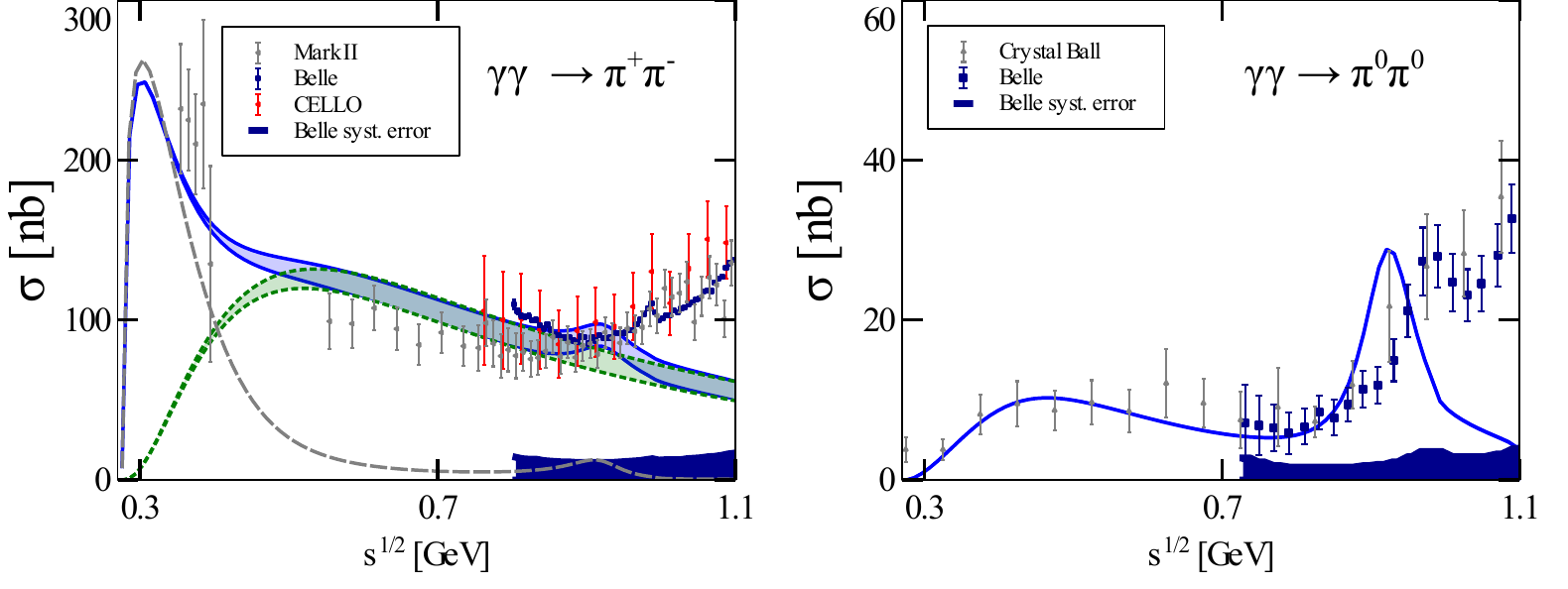}
  \caption{Cross section for the reactions $\gamma\gamma\rightarrow \pi^+\pi^-$ (left) and
    $\gamma\gamma\rightarrow \pi^0\pi^0$ (right) using \eqref{Eq:Correlation of parameters},
    \eqref{Eq:NEW_parameters} together with $g_5\in[-5,5]$. The obtained region is limited by the full thick lines.
    For the charged pions (left) the S-wave (long-dashed) and D-wave (short-dashed) are shown separately.
    See the figure caption of Fig.\ \ref{Fig:Cross_sections_GGtoPiPi_different_g3g5hO} for more details.}
  \label{Fig:Cross_sections_GGtoPiPi_final}
\end{figure*}
\begin{figure*}
  \centering
  \includegraphics[keepaspectratio,width=\textwidth]{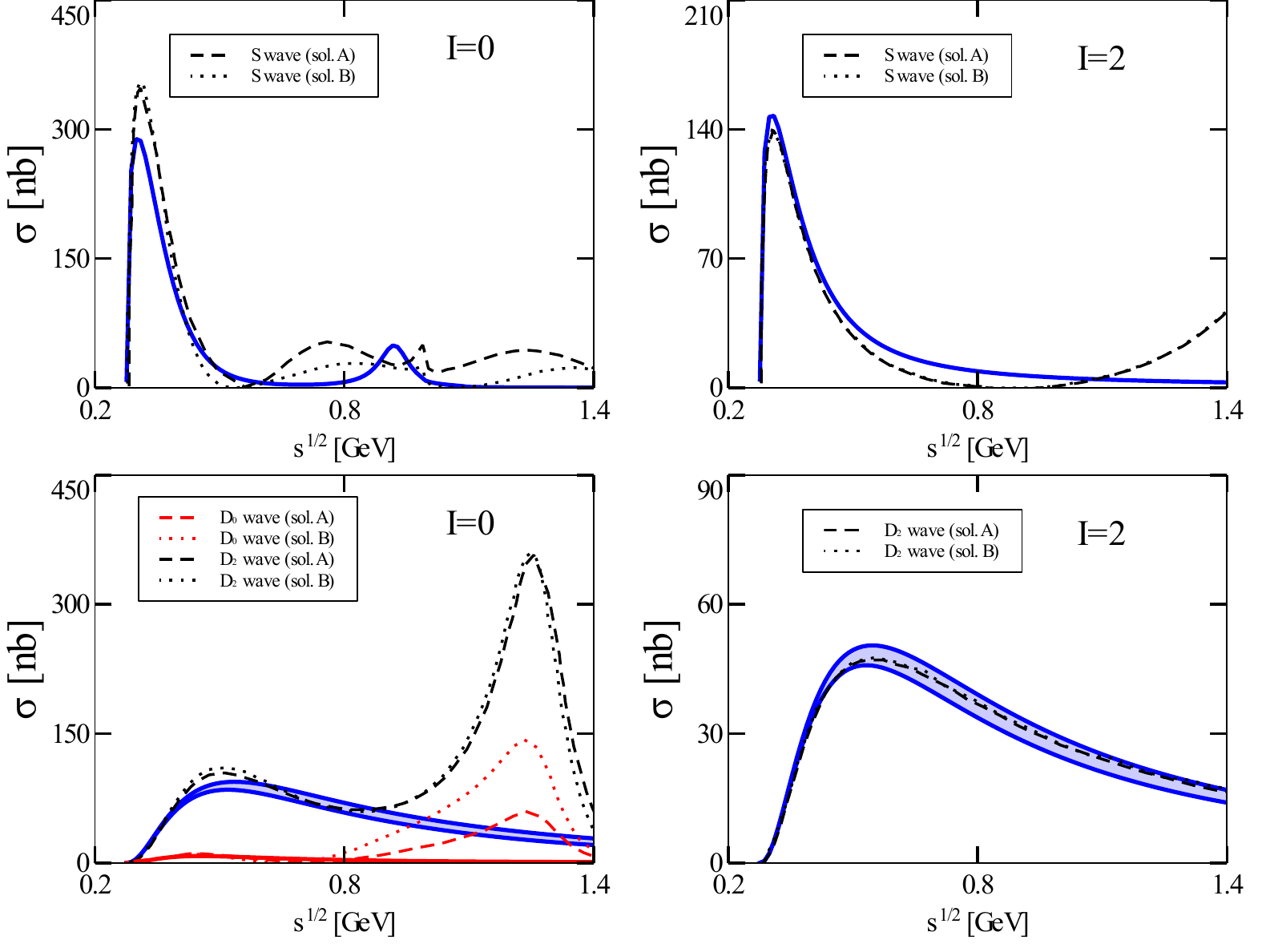}
  \caption{Comparison of our results (solid lines) with the
    results of \cite{Pennington:2008xd} for S-wave (top panels) and D-waves (bottom panels) and for different
    isospin (left panels: $I=0$; right panels: $I=2$). The subscript for the D-waves denotes the helicity. According to
    \cite{Pennington:2008xd} two solutions have been obtained:
    solution A (dashed) is favored by a $\chi^2$ fit, while solution B (dotted) is at the edge of acceptability.}
  \label{fig:pennington}
\end{figure*}
The cross sections for the two-pion channels are depicted in Fig.\
\ref{Fig:Cross_sections_GGtoPiPi_final}. Obviously both channels
$\pi^+\pi^-$ and $\pi^0\pi^0$ are well described up to energies of
about $\sqrt{s} \approx 0.9\,$GeV. Then our calculations show a
distinct peak, most pronounced in the neutral channel. After this
peak our theory curves decrease while the data continue to rise.
Two issues need to be disentangled here, namely the location of
the $f_0(980)$ in the S-wave and the rise towards the tensor
mesons in the D-wave. To do this we compare our results also to
the partial-wave analysis of \cite{Pennington:2008xd} as shown in
Fig.\ \ref{fig:pennington}. 

For the D-waves (bottom panels) we
observe reasonable agreement up the point in energy where the peak
from the isoscalar tensor meson starts out. As already stressed in
the introduction we expect that in the spirit of the hadrogenesis
conjecture this peak will be generated by vector-vector channels.
But since this is beyond the present work we cannot expect to
obtain a reasonable description of the D-wave beyond about 0.9
GeV. Below this energy the agreement is very satisfying. 

Turning
to the S-wave we observe also good agreement for isospin $I=2$
(top right panel in Fig.\ \ref{fig:pennington}). For the isoscalar
channel (top left) some disagreement with the results of
\cite{Pennington:2008xd} is observed. Most notably our peak for the $f_0(980)$ is slightly shifted to
lower energies, i.e.\ this dynamically generated scalar-isoscalar
state is somewhat overbound in our approach. This has already been
observed in \cite{Danilkin:2011fz}. Whether this is due to
higher-order effects in the scattering kernel or due to missing
vector-vector channels remains to be seen. We point the reader to the
fact that the leading order analysis \cite{Danilkin:2011fz} involved
two known parameters only, the chiral limit value of the pion decay
constant  and the coupling characterizing the decay of
the rho meson into a pair of pions. Therefore we do not have a
free parameter here to tune the location of the $f_0$ resonance.

A more precise description of the $f_0$ resonance is currently achieved
using Roy-Steiner equations  \cite{Ananthanarayan:2000ht,GarciaMartin:2011jx}
that are constrained by high-energy data. In particular in \cite{Moussallam:2011zg}
experimental input below the kaon-threshold is used in order to
constrain the shape of the inelasticity and the phase shift at the
$K\bar{K}$ threshold. The solutions of the dispersion integrals
yield a $f_0$ resonance and accurately predict its associated pole in the complex plane.

\begin{figure*}[t]
  \centering
  \includegraphics[keepaspectratio,width=\textwidth]{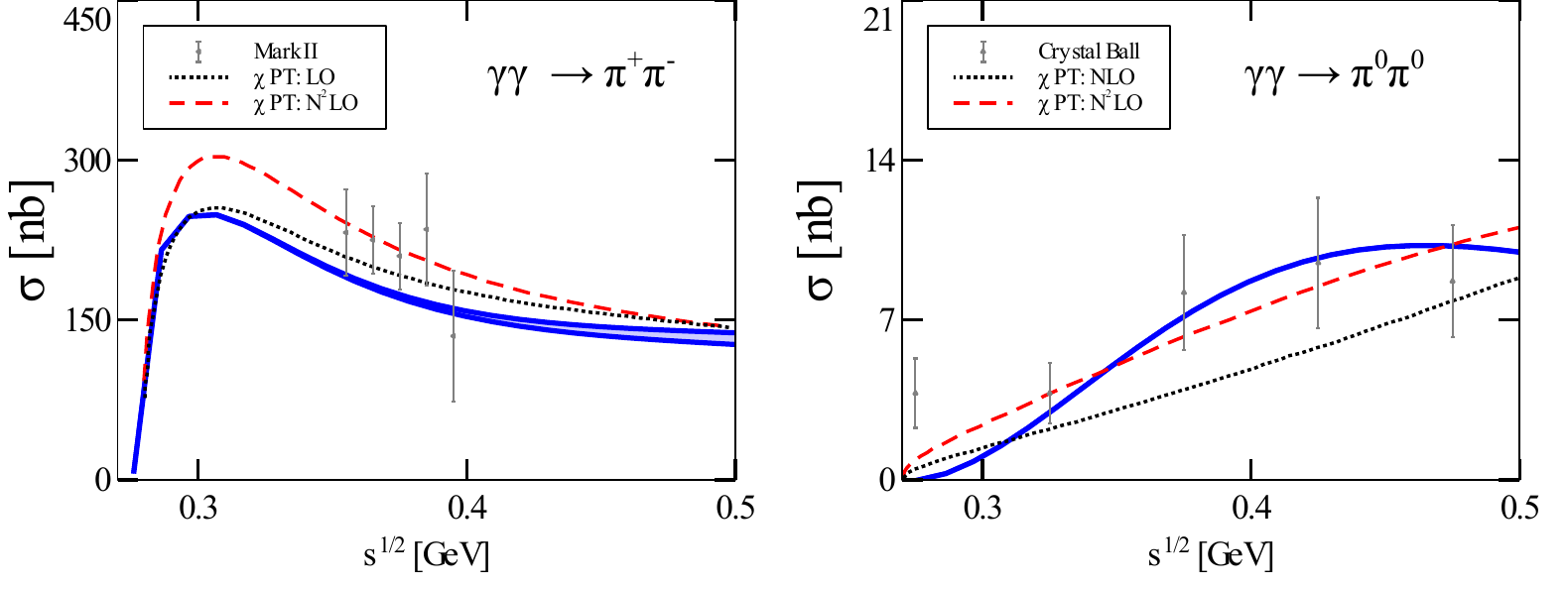}
  \caption{Comparison of our results (full lines) to the calculations from chiral perturbation theory ($\chi$PT) and to data.
    The dashed lines denote the next-to-next-to-leading-order calculations. The dotted lines denote the respective
    lowest-order non-trivial $\chi$PT result, which is leading order for the charged case and next-to-leading order for the neutral case.
    See also the figure caption of Fig.\ \ref{Fig:Cross_sections_GGtoPiPi_different_g3g5hO} for more details.}
  \label{fig:chiptcomp}
\end{figure*}
\begin{figure*}[t]
  \centering
  \includegraphics[keepaspectratio,width=\textwidth]{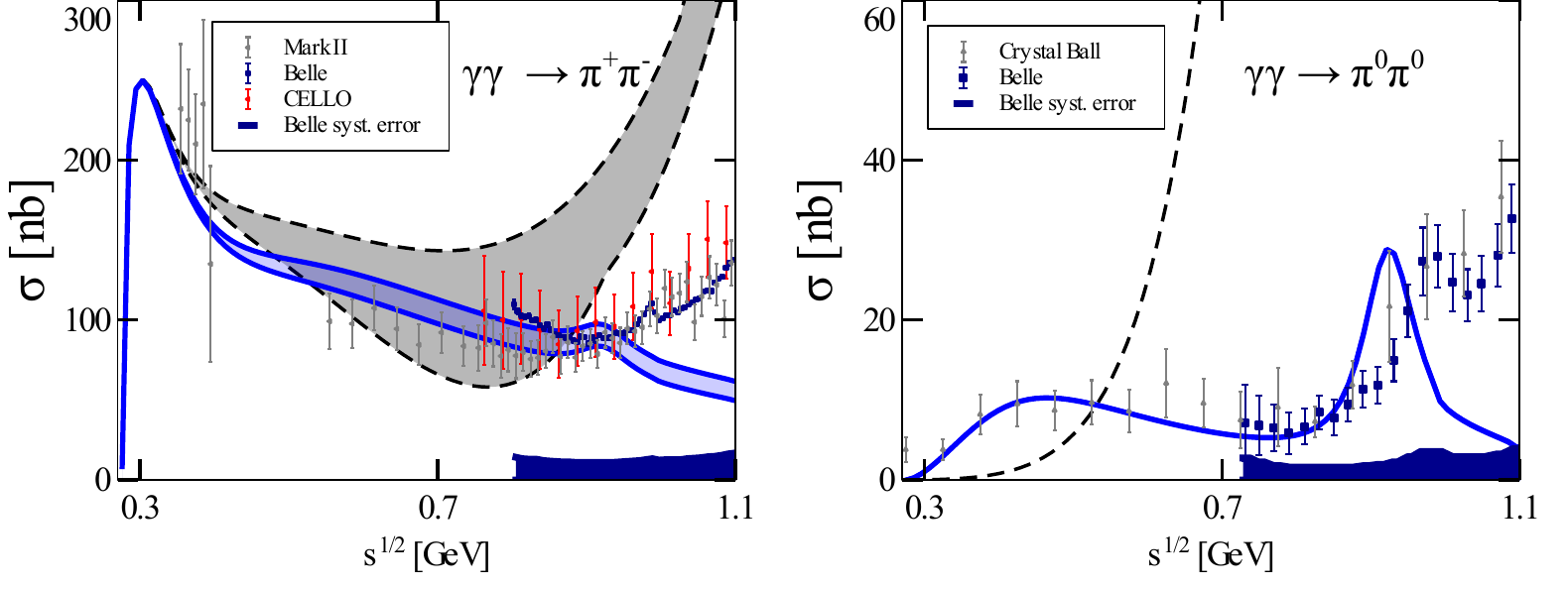}
  \caption{Comparison of tree-level calculations to the full results and to
  data. Tree-level calculations are depicted by dashed lines. See also the figure caption of Fig.\ \ref{Fig:Cross_sections_GGtoPiPi_different_g3g5hO} for more details.
    }
  \label{fig:tree-level-pi}
\end{figure*}
%

%%%%%%%%%%%%%%%%%%%%%%%%%%%%%%%%%%%%%%%%%%%%%%%%%%%%%%%%%%%%%%%%%%%%%%%%%%%%%%%%%%%%%%%%%%%
% Comment on dispersive analysis, ....

%%%%%%%%%%%%%%%%%%%%%%%%%%%%%%%%%%%%%%%%%%%%%%%%%%%%%%%%%%%%%%%%%%%%%%%%%%%%%%%%%%%%%%%%%%%

For the energy range below 0.9 GeV we deduce from Fig.\
\ref{Fig:Cross_sections_GGtoPiPi_final} that we have obtained an
overall good description of the reaction data. This is completely
in line with the complementary information contained in the pion
phase shifts as addressed in \cite{Danilkin:2011fz}. In Fig.\
\ref{fig:chiptcomp} we compare our calculations to the results
from $\chi$PT
\cite{Bijnens:1987dc,Donoghue:1988eea,Gasser:2005ud,Gasser:2006qa}.
We observe satisfying agreement. Note that even without vector
mesons our calculations contain multi-loop diagrams by the
achieved resummation in the $s$-channel. On the other hand, our
calculation does not contain all one-loop diagrams in the $t$- and
$u$-channel which enter $\chi$PT at next-to-leading order. In view
of these differences one can be satisfied with the agreement and
conclude that the numerically most important corrections from the
$\chi$PT point of view are included in our approach. At larger
energies pure $\chi$PT ceases to work and resummations must be
incorporated in one or the other way
\cite{Oller:1997ti,Oller:1997yg,GomezNicola:2001as,Danilkin:2011fz,Danilkin:2012ap}.

Finally we show in Fig.\ \ref{fig:tree-level-pi} the result of a
tree-level calculation based on our amplitudes \eqref{Eq:F1F2}.
Obviously the charged-pion channel is fairly insensitive to
rescattering effects, i.e.\ to a large extent dominated by the
one-pion exchange, which is responsible for the steep rise of the
cross section at low energies. The neutral-pion channel, however,
which does not have the corresponding one-pion exchange, is
dominated by loop/rescattering effects. In this channel the
tree-level calculation fails already at low energies.

\begin{figure*}[t]
\begin{center}
\vskip-0.1cm \hskip-0.5cm
\parbox{17.0cm}{\includegraphics[keepaspectratio,width=\textwidth]{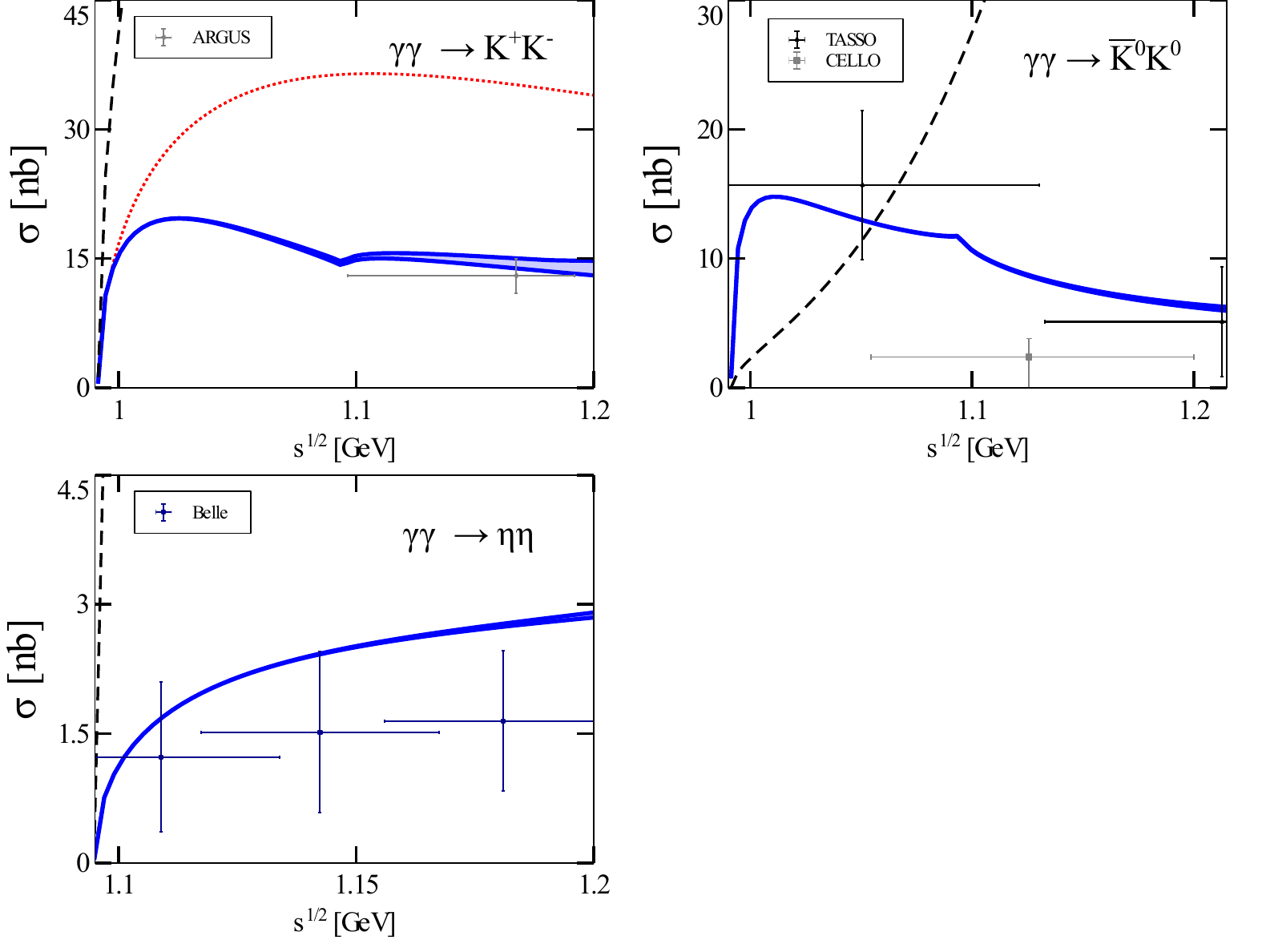}}
\parbox{13.5cm}{\vspace{-9cm}
\hspace*{7.3cm}
\parbox{7.5cm}{\caption{\label{fig:etakaon}
    Cross sections for the reactions $\gamma\gamma\rightarrow K^+K^-$ (top left),
    $\gamma\gamma\rightarrow K^0 \bar K^0$ (top right) and $\gamma\gamma\rightarrow \eta\eta$ (bottom left) using
    \eqref{Eq:Correlation of parameters}, \eqref{Eq:NEW_parameters} together with $g_5\in[-5,5]$. The obtained region is
    limited by the full lines. Tree-level calculations are depicted by dashed lines. For the charged-kaon case a pure Born-term
    calculation (one-kaon exchange) is shown by the dotted curve. The data are taken from
    \cite{Albrecht:1989re,Behrend:1988hw,Althoff:1985yh,Uehara:2010mq}.
    }}}
\end{center}
\end{figure*}

While the previously discussed channels have been used to some
extent to fix our free parameters, the channels which we discuss
in the following are pure predictions. Unfortunately the data
situation is rather poor in all three channels
$\gamma\gamma\rightarrow K^+K^-$, $K^0\bar{K}^0$ and $\eta\eta$,
but we will see that it is a non-trivial task to match the
available data points. We restrict ourselves to the energy region
close to threshold, i.e.\ to $\sqrt{s} \le 1.2\,$GeV. There we
expect the S-wave to dominate such that we do need to worry about
the tensor mesons.

The reaction $\gamma\gamma\rightarrow K^+K^-$ is depicted in the
top left panel of Fig.\ \ref{fig:etakaon}. Other theory approaches
have been reported in
\cite{Oller:1997yg,Achasov:2009ee,Lee:1998mz}. Unfortunately there
is only one data point with a large energy uncertainty in the
considered energy interval. Nonetheless, this data point is
significantly lower than generic tree-level calculations. For
comparison we show two types of such tree-level calculations. The
dashed line is obtained if our Lagrangian is used directly for the
amplitude and not for the potential of the full coupled-channel
calculation. An alternative tree-level approach is to use just the
kaon-exchange Born diagrams. We recall that the corresponding
pion-exchange Born diagrams are very significant for the
low-energy part of the reaction $\gamma\gamma\rightarrow
\pi^+\pi^-$. (This is the $\chi$PT-LO curve of Fig.\
\ref{fig:chiptcomp}.) For the kaon case the situation is obviously
different. While tree-level calculations fail to reproduce even
the close-to-threshold data, our full coupled-channel approach
leads to a significant reduction of the Born amplitude and matches
the available data point very nicely. Hence, the final-state
interactions are strong in this channel. A similar finding has
been reported in \cite{Oller:1997yg}. Finally we note that our
approach shows a visible cusp at the two-eta threshold. It is even
more pronounced in the neutral-kaon channel to which we turn next.

The reaction $\gamma\gamma\rightarrow K^0 \bar K^0$ is shown in
the top right panel of Fig.\ \ref{fig:etakaon}. The data, albeit
with large error bars, point to an initial steep increase of the
cross section with energy followed by a not so rapid fall. This
behavior is qualitatively reproduced by our full calculation,
though we do not fully match the second data point quantitatively.
Tree-level calculations cannot reproduce at all this rise-and-fall
behavior. Indeed, it is natural to expect that final-state
interactions are strong because the photons couple stronger to the
intermediate charged states than to the final neutral ones.

The bottom panel of Fig.\ \ref{fig:etakaon} shows the cross
section for the reaction $\gamma\gamma\rightarrow \eta\eta$ (see
also \cite{Lee:1998mz}). The data suggest a rather flat energy
dependence which cannot be reproduced by a pure tree-level
calculation (dashed line). In contrast, our full calculation (full
lines) including rescattering meets this requirement of a
comparatively flat cross section.
%To the best of our knowledge
%the only other theoretical analysis of the $\gamma\gamma\rightarrow \eta\eta$ process has been presented in \cite{Lee:1998mz}.
%There a very small cross section (below 0.1 nb) has been predicted.

We stress again that our results for the two-eta and two-kaon
channels are pure predictions. These channels did not enter the
determination of free parameters. Note also that the results are
basically insensitive to the remaining free parameter $g_5$.
Clearly, better data in these channels would be highly welcome to
further check the validity of our coupled-channel approach with
dynamical vector mesons.

\section{Summary and outlook}
\label{sec:sum}

We have performed a controlled study of the reactions
$\gamma\gamma\rightarrow \pi^0\pi^0$, $\pi^+\pi^-$,
$K^0\bar{K}^0$, $K^+K^-$, $\eta\,\eta$ and $\pi^0\eta$ in the
energy regime between the respective thresholds and about 1.2 GeV.
The reaction amplitudes were derived from the chiral Lagrangian
with dynamical vector meson fields properly constrained by
maximal analyticity and coupled-channel unitarity.

There are 5 unknown parameters, which have been constrained from
the reactions $\gamma\gamma\rightarrow \pi^0\pi^0$, $\pi^+\pi^-$,
$\pi^0\eta$ and from the differential decay $\eta \to \pi^0 \gamma
\gamma$. In particular we have achieved an excellent description
of the reaction $\gamma\gamma\rightarrow \pi^0\eta$ with its
lowest-lying scalar-isovector $a_0(980)$ resonance. The $a_0(980)$
resonance peak position does not depend on any of the free
parameters. Based on our parameter constraints we predict the
low-energy  $\gamma\gamma\rightarrow K^+K^-$, $K^0\bar{K}^0$ and
$\eta\,\eta$ cross sections.

While the vector mesons do play a crucial role in the derivation
of the generalized potentials for $\gamma\gamma\to P\,P $ and
$P\,P\to P\,P$ with $P=\pi,K,\eta $, the feedback of $P\,P\to
P\,V, \,V\,V$ reactions with $V = \rho,\omega, K^*, \phi$ remains
to be studied systematically. According to the hadrogenesis
conjecture we expect a quantitative description of the reactions
$\gamma\gamma\rightarrow \pi^0\pi^0$, $\pi^+\pi^-$,
$K^0\bar{K}^0$, $K^+K^-$, $\eta\,\eta$ and $\pi^0\eta$ up to about
2 GeV once such channels are incorporated in a controlled manner.

Accurate low-energy photon fusion data in particular in the
strangeness channels would further scrutinize the intricate
three-flavour dynamics of the Goldstone bosons and light vector
mesons.

\begin{acknowledgements}
The work of CT and SL has been supported by the European Community
Research Infrastructure Integrating Activity ``Study of Strongly
Interacting Matter'' (HadronPhysics3, Grant Agreement No.\ 283286)
under the Seventh Framework Programme of the EU.
\end{acknowledgements}

% BibTeX users please use one of
%\bibliographystyle{spbasic}      % basic style, author-year citations
%\bibliographystyle{spmpsci}      % mathematics and physical sciences
%\bibliographystyle{spphys}       % APS-like style for physics
%\bibliography{}   % name your BibTeX data base

\bibliographystyle{spphys}
\bibliography{GammaGamma_revised}{}

%% For one-column wide figures use
%\begin{figure}
%% Use the relevant command to insert your figure file.
%% For example, with the graphicx package use
%  \includegraphics{example.eps}
%% figure caption is below the figure
%\caption{Please write your figure caption here}
%\label{fig:1}       % Give a unique label
%\end{figure}
%%
%% For two-column wide figures use
%\begin{figure*}
%% Use the relevant command to insert your figure file.
%% For example, with the graphicx package use
%  \includegraphics[width=0.75\textwidth]{example.eps}
%% figure caption is below the figure
%\caption{Please write your figure caption here}
%\label{fig:2}       % Give a unique label
%\end{figure*}

\end{document}